\makeatletter
\def\input@path{{sn-article-template/}}
\makeatother
\documentclass[pdflatex,sn-mathphys-num]{sn-jnl}

\usepackage{xcolor}
\usepackage{graphicx}
\usepackage{amsmath}

\usepackage{tikz}
\usetikzlibrary{positioning,arrows.meta, matrix, arrows, calc, patterns.meta, hobby}

\definecolor{ObservedColor}{HTML}{FF7F00}
\colorlet{ObservedFill}{ObservedColor!70!white}

\definecolor{SimulatedColor}{HTML}{377EB8}
\colorlet{SimulatedFill}{SimulatedColor!70!white}

\definecolor{OverlapColor}{HTML}{999999}
\colorlet{OverlapFill}{OverlapColor!70!white}

\def\ObsCx{1.55}
\def\ObsCy{2.30}
\def\ObsRx{1.25}
\def\ObsRy{0.65}
\def\ObsRot{20}

\def\SimCx{2.45}
\def\SimCy{1.70}
\def\SimRx{0.95}
\def\SimRy{1.15}
\def\SimRot{-10}

\pgfmathsetmacro{\ObsXHalf}{sqrt((\ObsRx*cos(\ObsRot))^2 + (\ObsRy*sin(\ObsRot))^2)}
\pgfmathsetmacro{\ObsYHalf}{sqrt((\ObsRx*sin(\ObsRot))^2 + (\ObsRy*cos(\ObsRot))^2)}
\pgfmathsetmacro{\ObsXmin}{\ObsCx - \ObsXHalf}
\pgfmathsetmacro{\ObsXmax}{\ObsCx + \ObsXHalf}
\pgfmathsetmacro{\ObsYmin}{\ObsCy - \ObsYHalf}
\pgfmathsetmacro{\ObsYmax}{\ObsCy + \ObsYHalf}

\pgfmathsetmacro{\SimXHalf}{sqrt((\SimRx*cos(\SimRot))^2 + (\SimRy*sin(\SimRot))^2)}
\pgfmathsetmacro{\SimYHalf}{sqrt((\SimRx*sin(\SimRot))^2 + (\SimRy*cos(\SimRot))^2)}
\pgfmathsetmacro{\SimXmin}{\SimCx - \SimXHalf}
\pgfmathsetmacro{\SimXmax}{\SimCx + \SimXHalf}
\pgfmathsetmacro{\SimYmin}{\SimCy - \SimYHalf}
\pgfmathsetmacro{\SimYmax}{\SimCy + \SimYHalf}

\tikzset{
    domain box/.style={
        draw=black,
        thick
    },
    plume/.style={
        very thick
    },
    observed/.style={
        plume,
        draw=ObservedColor
    },
    simulated/.style={
        plume,
        draw=SimulatedColor
    },
    overlap fill/.style={
        fill=OverlapColor
    },
    overlap dots/.style={
        pattern={
            Dots[
                distance=4pt,
                radius=0.9pt
            ]
        },
        pattern color=black
    },
    centroid/.style={
        circle,
        fill=black,
        inner sep=1.8pt
    },
    observed extent/.style={
        <->,
        >=Stealth,
        thick,
        draw=ObservedColor
    },
    simulated extent/.style={
        <->,
        >=Stealth,
        thick,
        draw=SimulatedColor
    },
    observed guide/.style={
        thin,
        dashed,
        draw=ObservedColor
    },
    simulated guide/.style={
        thin,
        dashed,
        draw=SimulatedColor
    }
}

\usepackage{caption}
\usepackage{subcaption}

\usepackage{float}

\newcommand{\COtwo}{\textnormal{CO\textsubscript{2}}}

\usepackage{cleveref}

\begin{document}

\title[Bayesian seismic inversion of multilayer \COtwo{} migration]{Bayesian inversion of multilayer \COtwo{} migration from seismic plume observations using a graph-based finite-rate invasion-percolation model}

\author*[1]{\fnm{Elling} \sur{Svee}}\email{elling.svee@ntnu.no}
\author[1]{\fnm{Jo} \sur{Eidsvik}}
\author[1]{\fnm{Kasper} \sur{Hunnestad}}
\author[1]{\fnm{Philip} \sur{Ringrose}}

\affil*[1]{\orgname{Norwegian University of Science and Technology},
  \orgaddress{\city{Trondheim}, \country{Norway}}}

\abstract{Vertical migration of \COtwo{} in layered sandstone reservoirs is controlled by thin shale barriers whose properties are often poorly known. We develop a Bayesian framework that uses time-lapse seismic plume observations to estimate effective parameters governing lateral and vertical \COtwo{} migration. The forward model is a fast graph-based invasion-percolation model that extends conventional IP by representing both capillary-controlled filling of structural traps beneath shale barriers and finite-rate transfer through them. Parameters are inferred with approximate Bayesian computation and sequential Monte Carlo sampling, and the posterior samples are propagated into forecasts. Applied to real data from Sleipner, posterior simulations reproduce the broad distribution of \COtwo{} across nine sand units and its redistribution between 2010 and 2023. In contrast, the quasi-static model fails to reproduce this temporal evolution. Complementary synthetic experiments assess parameter recovery, forecasting, monitoring duration, and model misspecification. These experiments show that the information gained from monitoring depends on the migration events captured, with breakthrough and post-breach redistribution providing particularly strong constraints. This combination of fast simulation, probabilistic updating, and interpretable effective parameters makes the framework well suited to repeated forecast revision during active injection, especially when full-physics inference is too computationally demanding.}


\maketitle

\let\WriteBookmarks\relax
\def\floatpagepagefraction{1}
\def\textpagefraction{.001}

\renewcommand{\topfraction}{0.9}       
\renewcommand{\bottomfraction}{0.9}    
\renewcommand{\textfraction}{0.1}      
\renewcommand{\floatpagefraction}{0.7} 
\setcounter{topnumber}{4}              
\setcounter{bottomnumber}{4}           
\setcounter{totalnumber}{8}            


\section{Introduction}\label{seq:introduction}

The continued use of fossil fuels has increased \COtwo{} emissions and remains a major driver of climate change. Carbon capture and storage (CCS) offers a means of reducing industrial emissions by capturing \COtwo{}, transporting it to a storage site, and injecting it into deep geological formations. The effectiveness and safety of CCS depends on the injected \COtwo{} remaining contained within the storage complex. However, inferring subsurface \COtwo{} migration can be challenging because subsurface properties are often poorly characterized, while time-lapse seismic data may be noisy or have limited resolution \citep{arts_monitoring_sleipner_2004, chadwick_4d_2004}. A probabilistic framework is therefore needed to quantify uncertainty in both the inferred subsurface properties and the resulting future plume evolution.

Thin intra-formational barriers, such as shales, are an important source of uncertainty in layered storage reservoirs. These barriers strongly influence the lateral and vertical migration of \COtwo{} between permeable sand units. Their capillary properties depend on pore structure, interfacial tension, and wettability, which are difficult to determine because direct subsurface measurements are sparse \citep{cavanagh_sleipner_2014, iglauer_wettability_2015}. Furthermore, they may contain localized migration pathways where the \COtwo{} flows more easily \citep{martinez_enhanced_2025}. As a result, the onset and rate of \COtwo{} migration between reservoir layers may remain uncertain even when the larger-scale reservoir geometry is well characterized.

These uncertainties can be reduced by conditioning flow models on monitoring observations through history matching and data assimilation \citep{oliver_inverse_2008, oliver_recent_2011, crain_integrated_2024, teng_likelihood_free_2025}. However, such approaches typically require many forward simulations, making computational cost a central challenge. Full-physics multiphase simulators capture the relevant flow processes, but can be too expensive to use within Monte Carlo-based inference. Surrogate, reduced-order, and multifidelity models therefore provide attractive alternatives by trading some physical detail for substantially lower computational cost \citep{wen_ufno_2022, wen_nested_2023, han_surrogate_2024, watson_multifidelity_2025}.

For buoyancy- and capillary-dominated migration away from the injection well, invasion percolation (IP) provides an appropriate reduced-physics representation of \COtwo{} migration. It represents migration through capillary invasion and buoyancy-driven filling and can reproduce the main plume pathways at much lower computational cost than full-physics simulation \citep{carruthers_secondary_1998, carruthers_aapgdatapages_2003}. It has already been used to investigate the influence of multilayer shale properties at Sleipner \citep{cavanagh_sleipner_2014, callioli_santi_invasion_2025}. Yet, conventional IP models are quasi-static and treat migration across a breached shale as instantaneous. They therefore cannot represent the gradual inter-layer transfer needed to explain the continued evolution of lower accumulations after \COtwo{} has reached overlying units \citep{neufeld_modelling_2009, boait_spatial_2012}.

To address the need for a realistic representation of \COtwo{} migration that can be used for inference and forecasting, this paper develops a fast graph-based, multilayer IP simulator. Spill-point analysis is used to reduce each reservoir surface to a network of connected traps and spill paths, which is computed once and subsequently reused throughout the parameter estimate. Finite-rate vertical migration across shales is controlled by effective threshold pressures and post-breakthrough mobility parameters, which govern the onset and rate of vertical transfer, respectively. These parameters are inferred from time-lapse seismic plume outlines using approximate Bayesian computation with sequential Monte Carlo sampling (ABC-SMC). Forecasts of future plume evolution are then generated by propagating the inferred parameter uncertainty through the simulator, giving a probabilistic representation of the plume distribution and its uncertainty.

The framework is evaluated using both real-world observations and a controlled synthetic model. The Sleipner \COtwo{} storage site serves as a test case for assessing whether the reduced model can be conditioned on monitoring data to reproduce the observed plume distribution. It also enables us to investigate how uncertainty in shale topography propagates through the model and affects the inferred plume evolution. Complementing the Sleipner case, we consider a synthetic four-layer reservoir for which the true parameters and topographies are known. This controlled setting allows us to systematically assess parameter recovery, forecast uncertainty, the influence of monitoring duration, and the consequences of model misspecification.

The paper makes three main contributions.
\begin{itemize}
\item An efficient graph-based formulation of multilayer IP simulator with finite-rate migration across intra-formational barriers. In the quasi-static limit, it is about $40$ times faster than a cell-by-cell implementation on the reported benchmark.
\item A likelihood-free Bayesian treatment of uncertain barrier properties using time-lapse seismic plume outlines as observations.
\item A systematic assessment of how well the inferred model predicts plume evolution under both real-world and synthetic conditions.
\end{itemize}

\section{Invasion-percolation model}\label{seq:model}

\subsection{Physical assumptions and capillary breakthrough}\label{seq:quasi_static_ip_model}

IP theory is used as a reduced-order model of far-field \COtwo{} migration, and represents migration as a sequence of capillary invasion and structural filling events \citep{wilkinson_invasion_percolation_1983, carruthers_secondary_1998, carruthers_aapgdatapages_2003}. By avoiding the solution of large systems of coupled nonlinear equations required by full-physics models, it enables efficient simulation of plume evolution. The approximation is appropriate when viscous pressure gradients are small relative to capillary forces. The relative importance of viscous and capillary forces is characterized by the capillary number
\[
\mathrm{Ca}=\frac{\mu U}{\gamma},
\]
where $\mu$ is a representative fluid viscosity, $U$ is the fluid velocity, and $\gamma$ is the \COtwo-brine interfacial tension. A limit of approximately $\mathrm{Ca} \approx 10^{-4}$ is commonly used to identify capillary-dominated migration \citep{cavanagh_sleipner_2014}. For the Sleipner reservoir, \citet{cavanagh_sleipner_2014} estimate a capillary number of about $10^{-7}$, which is well below the required threshold.

We consider $N$ permeable sand units $\mathcal{S}_1,\ldots,\mathcal{S}_N$, ordered from deepest to shallowest. Let $\mathcal{B}_i$ denote the shale barrier immediately above $\mathcal{S}_i$. For $i<N$, $\mathcal{B}_i$ separates $\mathcal{S}_i$ from $\mathcal{S}_{i+1}$. At the pore scale, the pressure required for a non-wetting fluid to enter a pore throat is
\[
P_{\mathrm{th}}(r)=\frac{2\gamma\cos\xi}{r},
\]
where $\xi$ is the contact angle measured through the brine and $r$ is the pore-throat radius. For a water-wet system, the smaller pore throats of shale result in a higher entry pressure than in the surrounding sand \citep{berg_capillary_1975, espinoza_caprock_2017}. Migration between sand units is controlled by the capillary entry pressure of these barriers.

In the reduced model, each barrier $\mathcal{B}_{i}$ is assigned a single effective threshold pressure $P_{\mathrm{th},i}$. \COtwo{} accumulates beneath the barrier until the buoyancy pressure at the top of the accumulation exceeds this threshold. Within each connected accumulation, we assume vertical pressure equilibrium and a horizontal gas-water contact. Let $h_{i,j,t}$ denote the local \COtwo{} column height in horizontal grid cell $j$ beneath barrier $\mathcal{B}_{i}$ at time $t$. The corresponding buoyancy pressure at the top of the column is $\Delta\rho_i g h_{i,j,t}$, where $g$ is gravitational acceleration and $\Delta\rho_i$ is the density difference between brine and \COtwo{} in layer $i$. Capillary breakthrough becomes possible when $\Delta\rho_i g h_{i,j,t} \geq P_{\mathrm{th},i}$, which defines the critical column height
\[
h_{\mathrm{crit},i}
=\frac{P_{\mathrm{th},i}}{\Delta\rho_i g}.
\]
As injection proceeds, \COtwo{} therefore accumulates within a sand unit according to its geometry until the critical column height is reached beneath the overlying barrier. Once breakthrough occurs, \COtwo{} migrates into the next sand unit, where a new accumulation develops until the next barrier is breached. Repetition of this process produces a sequence of \COtwo{} accumulations across the permeable units.

\subsection{Finite-rate transfer across shale barriers}\label{seq:finite_rate}

In conventional quasi-static IP, transfer through a breached barrier is assumed to be instantaneous. Subsequent \COtwo{} supply is therefore passed directly to the overlying unit, and the lower accumulation cannot grow beyond its critical column height. This approximation is appropriate when the timescale of vertical transfer is short relative to that of lateral migration \citep{carruthers_aapgdatapages_2003}. However, at reservoirs such as Sleipner, lower accumulations continue to evolve after vertical migration has begun \citep{neufeld_modelling_2009, boait_spatial_2012}. This behavior is inconsistent with instantaneous transfer and suggests that the vertical transfer timescale is not negligible. We therefore retain quasi-static redistribution within each sand unit, but introduce a finite transfer rate across the shale barriers.

Let $L_{i,j}$ be the thickness of $\mathcal{B}_{i}$ in cell $j$. We treat a breached part of the barrier as an effective Darcy-like connection, with flow driven by the buoyancy pressure in excess of the entry pressure. The volumetric transfer rate through cell $j$ is then
\begin{equation}
  Q_{i,j}(t)
  =\frac{\lambda A_j}{L_{i,j}}
   \left[\Delta\rho_{i} g h_{i,j}(t)-P_{\mathrm{th},i}\right]_+
  =\frac{\lambda\Delta\rho_{i} g A_j}{L_{i,j}}
   \left[h_{i,j}(t)-h_{\mathrm{crit},i}\right]_+,
  \label{eq:seal_flux}
\end{equation}
where $[x]+=\max(x,0)$ and $A_j$ is the horizontal area of cell $j$. We use one common effective mobility $\lambda$ with units $\mathrm{m^2,Pa^{-1},s^{-1}}$ for all intra-formational barriers, although the formulation could be extended to layer-specific mobilities. With this transfer law, the barrier threshold pressures and mobility together control the lateral and vertical distribution of \COtwo{} in the reservoir. The $P_{\mathrm{th},i}$ determines when upward migration begins from each unit, whereas $\lambda$ controls the subsequent rate. Both should be interpreted as effective model-scale parameters that incorporate the combined effects of the underlying pore-scale properties.

\Cref{eq:seal_flux} is a simplified representation of buoyancy-driven two-phase flow through a shale barrier. It neglects brine pressure, counter-current brine flow, and back-pressure from the receiving unit, all of which require more complete layered-aquifer models to represent explicitly \citep{jenkins2019gasinjectionleakagelayered}. Beyond these simplifications, the IP formulation describes only drainage, in which non-wetting \COtwo{} displaces the wetting brine phase. Like other IP simulators such as Permedia \citep{Halliburton_PermediaCO2_2025, ni_quantifying_2021}, it does not represent subsequent imbibition or residual trapping. Because the simulations considered here are restricted to the injection period, drainage is expected to dominate plume evolution, making imbibition and residual trapping of secondary importance. These processes could also be incorporated in future extensions of the model is desired.

\subsection{Graph-based multilayer simulation}

Repeated simulations of \COtwo{} migration require a computationally efficient forward model. The geometric structure of the migration problem allows us to construct such a model by reducing each shale topography to a graph of connected structural traps. Because this graph depends only on the reservoir geometry, it can be constructed once and reused across simulations. Rather than updating individual cells in a three-dimensional grid, the simulator routes \COtwo{} mass between these traps. For each surface, a spill-point analysis assigns cells to traps, identifies their outlets, and organizes nested and downstream traps into a directed hierarchy. For single units, such a construction is provided by \texttt{SurfaceWaterIntegratedModeling.jl} and described in detail by \citet{swim_package}. Its spill-point algorithms were originally used in \texttt{MRST-co2lab} to analyse \COtwo{} migration within a single sand unit \citep{moll_nilsen_spill_point_2015}. Here, we extend the approach to a multilayer setting, giving a graph-based representation of the entire reservoir.

\begin{figure}[H]
  \centering
  \begin{subfigure}[c]{0.46\linewidth}
    \centering
    \resizebox{\linewidth}{!}{\begin{tikzpicture}[
  x=1cm,
  y=1cm,
  use Hobby shortcut,
  tension=1.1,
  outline/.style={
    draw=black,
    line width=3pt,
    line cap=round,
    line join=round
  },
  migration/.style={
    draw=black,
    line width=3pt,
    dash pattern=on 5pt off 5pt,
    -{Latex[length=4mm,width=4mm]}
  },
  layerlabel/.style={
    font=\bfseries\Large,
    anchor=west
  },
  traplabel/.style={
    font=\bfseries\large,
    anchor=center
  },
  spillline/.style={
    draw=black,
    line width=1.5pt,
    dash pattern=on 4pt off 4pt
  }
]

\definecolor{layerblue}{RGB}{103,170,207}


\def\uppercurve{
  (0.0,3.65)
  .. (2.6,4.75)
  .. (4.4,4.3)
  .. (7.2,5.5)
  .. (10.4,4.8)
  .. (12.3,3.65)
}

\def\lowercurve{
  (0.00,0.00)
  .. (0.65,0.28)
  .. (1.35,0.90)
  .. (2.10,1.95)
  .. (2.75,2.38)
  .. (3.30,2.10)
  .. (3.85,1.72)
  .. (4.35,1.92)
  .. (5.05,2.75)
  .. (5.55,2.82)
  .. (6.15,2.20)
  .. (6.85,1.30)
  .. (7.65,0.72)
  .. (8.45,0.78)
  .. (9.25,1.28)
  .. (9.85,1.40)
  .. (10.45,1.08)
  .. (11.10,0.48)
  .. (12.30,0.00)
}


\node[layerlabel] at (0.12,5.72) {Storage layer 2};
\node[layerlabel] at (0.02,2.96) {Storage layer 1};


\begin{scope}
  \clip (4.35,4.55) rectangle (11.15,6.0);

  \path[fill=layerblue]
    \uppercurve
    -- (12.30,3.65)
    -- (0.00,3.65)
    -- cycle;
\end{scope}

\begin{scope}
  \clip
    \uppercurve
    -- (12.30,3.65)
    -- (0.00,3.65)
    -- cycle;

  \draw[spillline]
    (0.0,4.30) -- (11.20,4.30);
\end{scope}

\draw[outline] \uppercurve;
\draw[outline] (0.00,3.65) -- (12.30,3.65);


\begin{scope}
  \clip (0,0.35) rectangle (12.4,3.2);

  \path[fill=layerblue]
    \lowercurve
    -- (12.30,0.00)
    -- (0.00,0.00)
    -- cycle;
\end{scope}

\begin{scope}
  \clip
    \lowercurve
    -- (12.30,0.00)
    -- (0.00,0.00)
    -- cycle;

  \draw[spillline]
    (1.95,1.72) -- (6.55,1.72);

  \draw[spillline]
    (1.15,0.72) -- (10.85,0.72);
\end{scope}

\draw[outline] \lowercurve;
\draw[outline] (0.00,0.00) -- (12.30,0.00);

\draw[migration]
  (5.4,2.82)
  to[out=75,in=240]
  (7.80,5.6);


\node[traplabel] at (2.15,4.55) {A};
\node[traplabel] at (7.95,4.95) {B};
\node[traplabel] at (10.0,4.0) {C};

\node[traplabel] at (2.75,2.05) {A};
\node[traplabel] at (5.40,2.28) {B};
\node[traplabel] at (4.25,1.2) {C};
\node[traplabel] at (9.75,1.10) {D};
\node[traplabel] at (6.20,0.35) {E};

\end{tikzpicture}}
    \caption{Layered reservoir with \COtwo{} accumulations}
  \end{subfigure}
  \hspace{0.04\linewidth}
  \begin{subfigure}[c]{0.32\linewidth}
    \centering
    \resizebox{\linewidth}{!}{\begin{tikzpicture}[
  x=1cm,
  y=1cm,
  >=Stealth,
  node circle/.style={
    circle,
    draw=black,
    line width=3pt,
    minimum size=10mm,
    inner sep=0pt,
    font=\bfseries\large
  },
  edge/.style={
    draw=black,
    line width=3pt,
    -{Latex[length=4mm,width=4mm]}
  },
  migration/.style={
    draw=black,
    line width=3pt,
    dash pattern=on 5pt off 5pt,
    -{Latex[length=4mm,width=4mm]}
  }
]

\node[anchor=west,font=\bfseries\large]
  at (0,9.65) {Storage layer 2};

\node[anchor=west,font=\bfseries\large]
  at (0,5.20) {Storage layer 1};

\node[node circle] (u1) at (0.5,8.6) {A};
\node[node circle] (u3) at (7.65,8.45) {B};
\node[node circle] (u2) at (5.1,6.55) {C};

\draw[edge] (u1) -- (u2);
\draw[edge] (u3) -- (u2);

\node[node circle] (l1) at (0.5,4.1) {A};
\node[node circle] (l3) at (3.95,4.1) {B};
\node[node circle] (l2) at (2.15,2.45) {C};
\node[node circle] (l5) at (7.65,2.45) {D};
\node[node circle] (l4) at (5.1,0.55) {E};

\draw[edge] (l1) -- (l2);
\draw[edge] (l3) -- (l2);
\draw[edge] (l2) -- (l4);
\draw[edge] (l5) -- (l4);

\draw[migration]
  (l3.18)
  .. controls (5.65,4.65) and (7.30,5.65) ..
  (u3.265);

\node[
  anchor=west,
  align=center,
  font=\bfseries\large
] at (7.15,5.55)
  {Vertical\\migration};

\end{tikzpicture}}
    \caption{Graph representation}
  \end{subfigure}
  \caption{Cross-sectional (a) and Graph-representation (b) of a structural trap filling sequence. Nodes represent structural traps, and are labeled for clarity. Note that the directed edges do not represent the spill paths, but illustrate the hierarchy where traps combine into larger accumulations.}
  \label{fig:ip_illustration}
\end{figure}
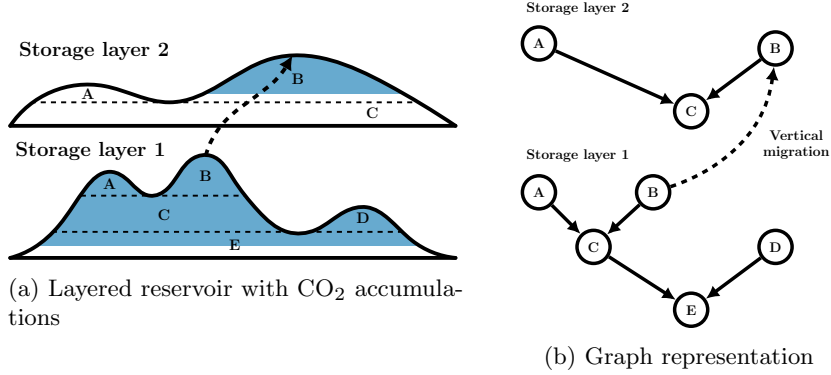

We construct directed graph for each sand unit, with structural traps represented as nodes and lateral spill connections as edges. Closed lateral boundaries are imposed, so \COtwo{} cannot spill out of the model domain. \Cref{fig:ip_illustration} illustrates the resulting hierarchy for a two-layer reservoir. Injected \COtwo{} is routed to a trap, where it accumulates until either the trap reaches its lateral spill point or the pressure required to breach the overlying shale is exceeded. In the former case, excess mass follows the corresponding edge to a downstream trap within the same graph. In the latter, \COtwo{} is transferred vertically to a directed graph for the above unit and  introduced as a source at the horizontal location of the breach. By processing the layers from deepest to shallowest, the simulator propagates mass through the coupled graphs. In the quasi-static limit, the simulator can then be used to compute the entire time evolution of the plume in a single pass. For finite-rate transfer, the simulator is called repeatedly at adaptive time intervals to resolve the vertical transfer.

The computational kernel is implemented in Rust \citep{rustlang}, with Python \citep{python_programming_language} bindings used for the inference workflow. Its efficiency in the quasi-static limit is assessed against a cell-by-cell implementation. We use the Sleipner reservoir, represented by nine shale surfaces on a $65 \times 119$ grid from the Sleipner 2019 Benchmark Model \citep{equinor2020sleipner}. The cell-by-cell reference discretizes the depth axis into 324 vertical layers. Runtime measurements over a $27$-year forward simulation were performed on a single core of an AMD Ryzen 7 5800X over 30 repetitions. The graph-based implementation achieved a median runtime of $7.09\,\mathrm{ms}$, whereas the cell-by-cell reference required $288\,\mathrm{ms}$, corresponding to an approximately $40$ times speedup. We separately assess the computational cost introduced by finite-rate vertical migration within the graph-based model. Finite-rate migration is more expensive because vertical transfer must be resolved through adaptive time refinement. Across $17$ values spanning $-14\leq\log_{10}\lambda\leq-4$, the median runtime was $64.2\,\mathrm{ms}$, about $9.1$ times that of the quasi-static graph-based model.

\section{Bayesian inversion from plume observations}\label{seq:inference}

\subsection{Observation model for plume outlines}\label{seq:seismic_detection}

\begin{figure}[htb]
  \centering
  \begin{subfigure}[t]{\linewidth}
    \centering
    \includegraphics[width=0.4\linewidth]{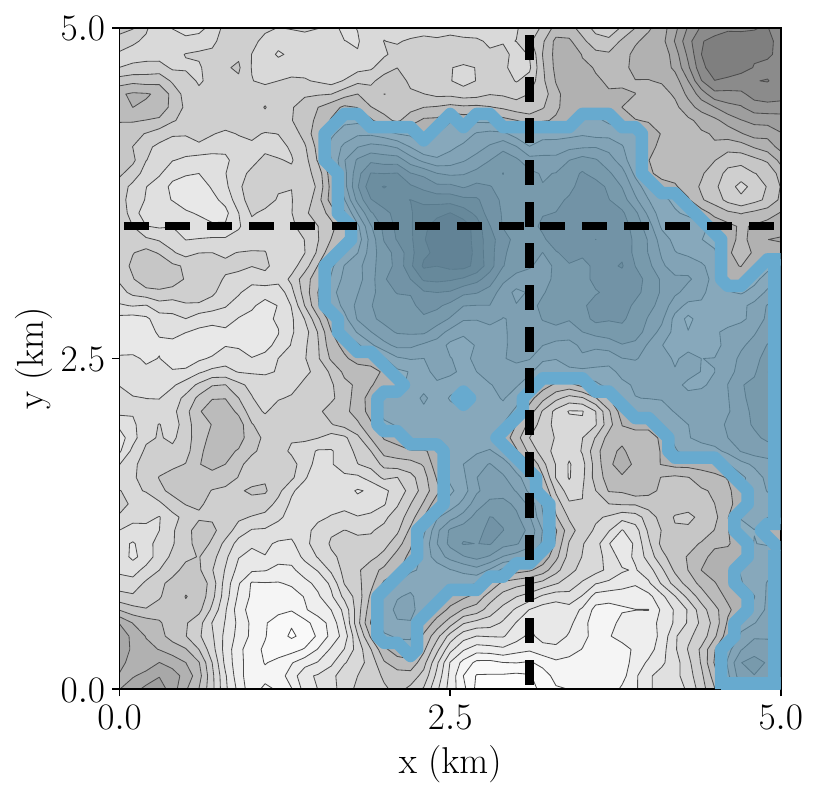}
    \caption{Bird's-eye view with the plume outline coloured blue and cross-sections shown as dashed black lines}
  \end{subfigure}
  \vfill
  \begin{subfigure}[t]{0.48\linewidth}
    \centering
    \includegraphics[width=\linewidth]{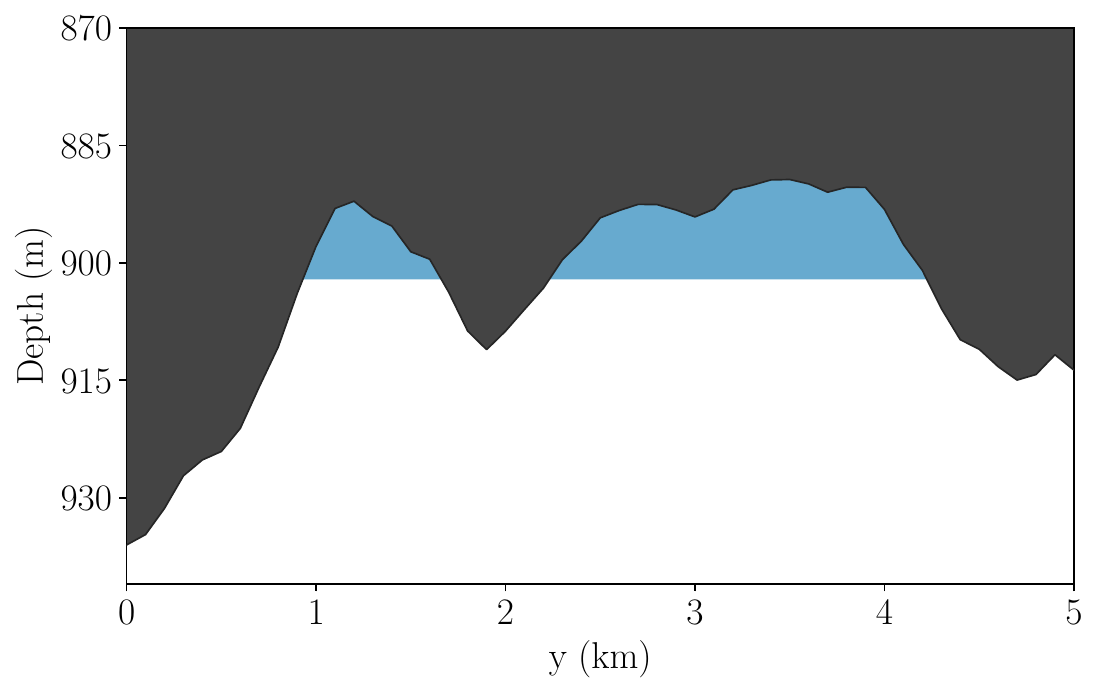}
    \caption{Cross-section along the $x$-axis}
  \end{subfigure}
  \hfill
  \begin{subfigure}[t]{0.48\linewidth}
    \centering
    \includegraphics[width=\linewidth]{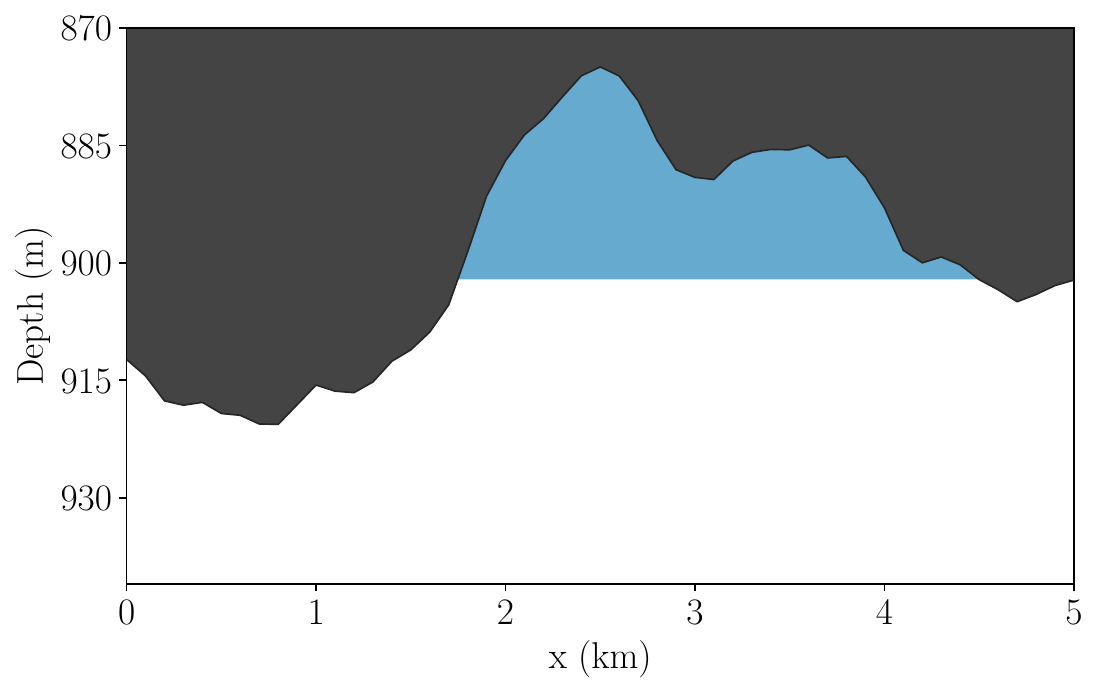}
    \caption{Cross-section along the $y$-axis}
  \end{subfigure}

  \caption{Illustration of the relation between the simulated column heights $h_{i,j,t}$ and the binary plume outlines $Y_{i,j,t}$}
  \label{fig:plume_outline}
\end{figure}

To model the uncertainty in the \COtwo{} migration, the parameters of the finite-rate IP model are inferred using Bayesian inversion. The inversion is informed by time-lapse seismic observations of the \COtwo{} plume. These surveys rely on amplitude-difference, travel-time, and waveform information to monitor plume evolution over time. As a simplified observation model, we represent the interpreted plume outlines from the seismic surveys as two-dimensional binary masks indicating whether \COtwo{} is detected in each horizontal cell. The relation between the simulated column heights and the binary outlines is illustrated in \Cref{fig:plume_outline}.

Recalling that $h_{i,j, t}$ denotes the local \COtwo{} column height in cell $j$ beneath barrier $\mathcal{B}_i$ at time $t$, we define the binary mask for sand unit $i$ as
\[
  Y_{i,j, t}
  =\mathbf{1}\!\left\{h_{i,j, t} \geq h_{\mathrm{det}}\right\},
\]
where $\mathbf{1}\{\cdot\}$ is the indicator function and $h_{\mathrm{det}} \geq 0$ is the minimum detectable column height. A non-zero $h_{\mathrm{det}}$ removes thin plume margins that might not be visible on limited-resolution seismic data. We mainly assume that $h_{\mathrm{det}}=0$ in this paper, corresponding to perfect detection of any non-zero column height.

We choose this observation model because it provides a simple and interpretable summary of plume evolution. However, it does not restrict the framework, as the simulator can be also combined with more complex observation models. An important limitation is that the column heights are derived from the topographic model, which may be uncertain or inaccurate in real applications. Errors in the topography can therefore introduce bias that is not accounted for in the current model.

\subsection{Likelihood-free parameter inference with ABC-SMC}

Using the simulator and observation model, we can relate the effective parameters of the finite-rate IP model to the observed plume outlines. We have $N$ parameters in total, $N-1$ threshold pressures and one shared mobility. Here the uppermost shale is assumed to be impermeable, although this could be relaxed. The inferred parameters are combined into a single vector
\[
  \boldsymbol{\theta}
  =(P_{\mathrm{th},1},\ldots,P_{\mathrm{th},N-1},\lambda).
\]
A prior distribution $\pi(\boldsymbol{\theta})$ is assigned to $\boldsymbol{\theta}$, representing the beliefs before conditioning on observations. The prior is application-dependent, but suitable choice is log-normal distributions to ensure positivity. At each survey time $t_k$, the observations provide one binary plume mask $\mathbf{Y}^{\mathrm{obs}}_{i,j, k}$ for sand unit $i$ and cell $j$. For a given parameter vector $\boldsymbol{\theta}$, the simulator and detection operator produce the corresponding simulated masks $\mathbf{Y}_{i,j,k}(\boldsymbol{\theta})$. We denote the complete collections of observed and simulated masks across all sand units and survey times by $\mathbf{Y}^{\mathrm{obs}}$ and $\mathbf{Y}(\boldsymbol{\theta})$, respectively.

In a Bayesian formulation, the likelihood $L(\mathbf{Y}^{\mathrm{obs}}\mid\boldsymbol{\theta})$ which describes the probability of the observations for a given parameter vector. Bayes' rule gives the posterior
\[
  \pi(\boldsymbol{\theta}\mid\mathbf{Y}^{\mathrm{obs}})
  \propto
  L(\mathbf{Y}^{\mathrm{obs}}\mid\boldsymbol{\theta})
  \pi(\boldsymbol{\theta}).
\]
However, for the interpreted plume outlines, no tractable likelihood is available. We therefore use ABC for likelihood-free inference \citep{beaumont_approximate_2002,csillery_approximate_2010}. ABC samples parameter values $\boldsymbol{\theta}_i$ from the prior, passes them through the simulator, and compares simulated output with the observations. This comparison is performed summary statistics \citep{sisson_handbook_abc_2018}, denoted by $\boldsymbol{s}(\boldsymbol{\theta}_i)$ and $\boldsymbol{s}^{\mathrm{obs}}$, respectively. By retaining parameter values that produce sufficiently close summaries, it approximates the posterior. However, directly sampling from the prior can be inefficient in high-dimensional parameter spaces. SMC improves this by evolving a population of parameter values through a sequence increasingly restrictive distributions toward the ABC posterior \citep{sisson_sequential_2007, toni_abc_smc_2009}.

ABC-SMC algorithm is implemented in \texttt{PyMC} \citep{salvatier_probabilistic_2015, castro_approximate_2025}. It has a Gaussian ABC kernel giving an approximate ABC posterior
\begin{equation}
  \pi_{\mathrm{ABC}}(\boldsymbol{\theta}\mid \boldsymbol{s}^{\mathrm{obs}})
  \propto
  \exp\!\left[-\frac{1}{2}d^2(\boldsymbol{\theta})\right]
  \pi(\boldsymbol{\theta}),
\end{equation}
where the discrepancy function $d(\boldsymbol{\theta})$ describes the distance between the observed and simulated summary vectors. For a set of $V$ summary statistics, the squared normalized discrepancy is defined as
\[
  d^2(\boldsymbol{\theta})
  =\sum_{v=1}^{V}
  \left(
    \frac{s_v(\boldsymbol{\theta})-s_{v}^{\mathrm{obs}}}{\epsilon_v}
  \right)^{2}.
\]
Here, $s_v(\boldsymbol{\theta})$ corresponds to the $v$-th component of the simulated summary vector, while $s_{v}^{\mathrm{obs}}$ is the corresponding observed summary. The $\epsilon_v$ are fixed componentwise discrepancy scales that set the relative influence of the summary components. Starting from an initial population drawn from the prior, SMC moves toward the posterior through a sequence of adaptively chosen tempering stages. At stage $w$, the population targets
\[
\pi_{\mathrm{ABC},w}(\boldsymbol{\theta}\mid \boldsymbol{s}^{\mathrm{obs}})
\propto
\pi(\boldsymbol{\theta})
\exp\!\left[-\frac{\beta_w}{2}d^2(\boldsymbol{\theta})\right],
\qquad 0=\beta_0<\beta_1<\cdots\leq 1.
\]
The next value of $\beta_w$ is chosen adaptively based on the effective sample size of the incremental importance weights, and the procedure continues until $\beta_w=1$. At each stage, the particles are reweighted and resampled, after which a Metropolis-Hastings mutation kernel is applied to restore diversity and explore the current target distribution. The \texttt{PyMC} implementation uses a multivariate normal proposal fitted to the particle population, with the number of mutation steps determined adaptively from the reduction in particle correlation.

\subsection{Plume-outline summary statistics}\label{seq:summary_statistics}

To ensure the ABC posterior is informative about the parameters of interest, the summary statistics should capture plume features related to both vertical migration between sand units and lateral migration within each unit. We use three summary statistics for each layer and observation time, with the summary vector expanded as additional surveys become available. The first statistic characterizes the distribution of \COtwo{} between the sand units, while the second and third statistics characterize the lateral position of the plume.

If the reservoir topography and plume outlines could be resolved with sufficient accuracy, the stored mass could be estimated by integrating plume thickness over the detected footprint. In practice, seismic imaging provides only an uncertain representation of the reservoir topography, which can bias such an estimate. We therefore use footprint area as a simpler proxy for the relative amount of \COtwo{} in each unit. Let $A_{i,k}=\sum_j Y_{i,j,k}$ denote the number of occupied cells in unit $i$ at observation time $k$, and let $a_k=\sum_{i=1}^{N}A_{i,k}$ be the total number of occupied cells across all units. For ease of notation, we omit denoting whether the mask $Y$ is observed or simulated. The first summary statistic is then
\[
  s_{1,i,k}
  =
  \frac{A_{i,k}}{\max\{a_k,1\}},
\]
and measures the fraction of the total detected footprint contained in unit $i$. For the second and third statistics, let $(x_j,y_j)$ denote the normalized coordinates of cell $j$. We define the area-weighted first moments
\begin{equation}\label{eq:summary_statistics_spatial_moments}
  s_{2,i,k}  = \frac{1}{\max \{a_{k},1\}}\sum_{j}x_j Y_{i,j,k},
  \qquad
  s_{3,i,k} = \frac{1}{\max \{a_{k},1\}}\sum_{j}y_j Y_{i,j,k}.
\end{equation}
Because these moments are normalized by the total detected footprint rather than the footprint within each unit, they capture both the lateral location of the plume and the relative amount present in that unit. If no plume is detected at time $k$, all three summaries are zero.

Restricting the summary vector to these three quantities balances retaining informative plume features against keeping the ABC comparison low-dimensional. Moreover, because the selected summaries are not known to be sufficient, the resulting ABC posterior depends on their choice and may differ from the posterior based on the full plume observations. Additional outline-based measures, such as intersection-over-union, could be included, but more summaries do not necessarily improve the inference. Redundant or weakly informative components can make simultaneous agreement with the observations more difficult and increase sensitivity to discrepancy scaling. 

\section{Application to the Sleipner storage site}\label{seq:sleipner}

\subsection{Dataset}

We now apply the framework to seismic plume outlines data from the Sleipner storage site in the North Sea. This provides a real-world test of the conditioning approach in a setting with a well-documented multilayer plume. Since the shale properties are unknown, the aim here is not to assess parameter recovery against a known truth. Instead, we examine whether the model reproduces the broad plume redistribution, and see which effective parameter combinations are informed by the data.

\begin{figure}[H]
  \centering
  \includegraphics[width=0.75\linewidth]{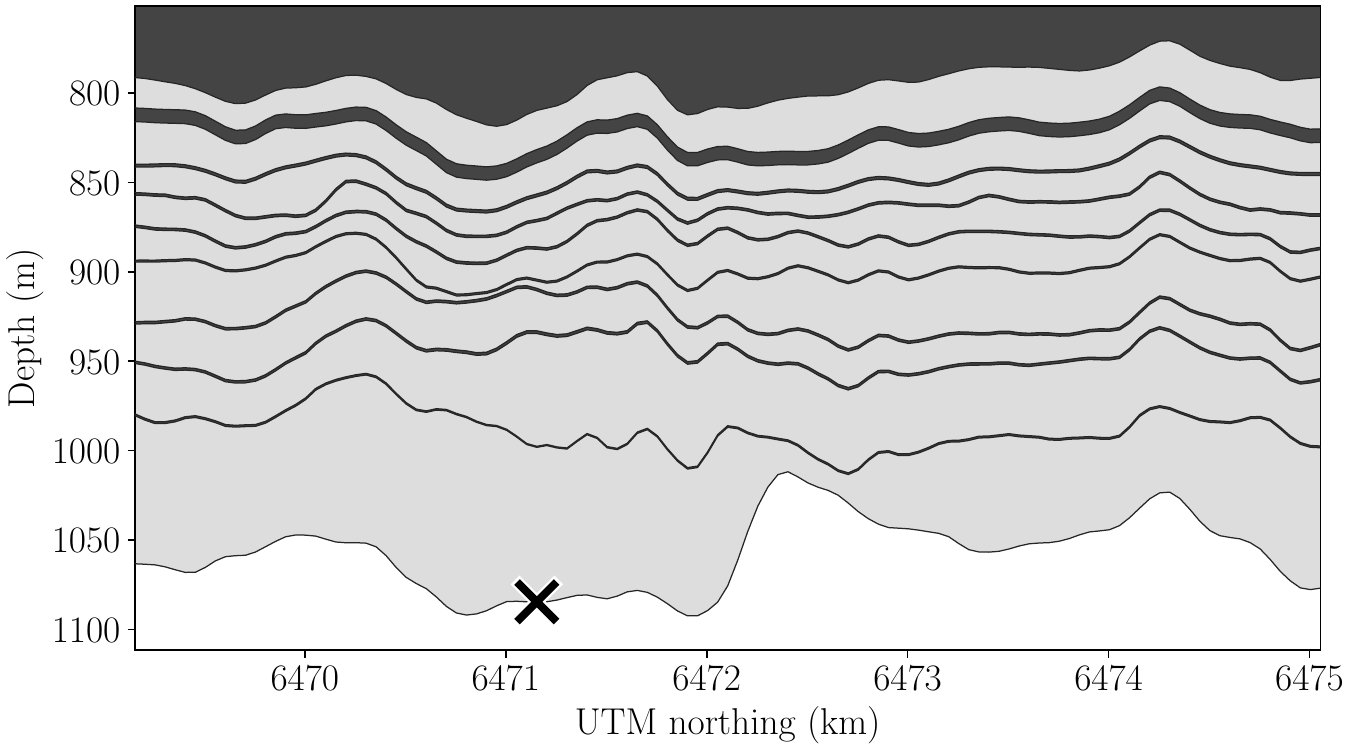}
  \caption{Cross-section through the Sleipner layers at a UTM easting of $438.5\,\mathrm{km}$. The cross marks the injection location.}
  \label{fig:SleipnerCrossSection}
\end{figure}

The storage site contains nine main sand units separated by eight intra-formational shale barriers. Seven barriers are approximately $1$-$2\,\mathrm{m}$ thick, while the uppermost has a mean modelled thickness of about $7.6\,\mathrm{m}$. A much thicker shale succession seals the formation above \citep{cavanagh_sleipner_2014}, and is treated as an impenetrable top barrier. \Cref{fig:SleipnerCrossSection} shows a cross-section through the shales at the injection location. Other thinner shale layers may occur within the main sand layers, but were not found to affect the \COtwo{} migration. Injection of \COtwo{} separated from natural-gas production began in $1996$ through a well near the base of the Utsira Formation. \citet[Table A]{callioli_santi_invasion_2025} summarizes the annual injection volumes. A baseline seismic survey was acquired in $1994$, followed by multiple repeat surveys between $1999$ and $2020$ \citep{arts_monitoring_sleipner_2004, ringrose_estimating_2026}. The plume reached the uppermost sand layer by the first repeat survey in $1999$ \citep{cavanagh_sleipner_2014}. This rapid ascent has been linked to localized feeder pathways through the intra-formational shales \citep{furre_observing_2019, martinez_unraveling_2026}. 

This study uses seismic monitoring data from the $2010$ and $2023$ surveys. The $2010$ outlines and shale surfaces come from the Sleipner $2019$ Benchmark Model in CO2DataShare \citep{equinor2020sleipner}, where the shale surfaces are represented on a $65 \times 119$ grid of $50\times50\,\mathrm{m}$ cells. The $2023$ outlines are derived from the full-waveform inversion results of \citet{martinez_unraveling_2026}. An important consideration is that the observed outlines are laterally mismatched with the topography, meaning that the outlined plume is not located in the structural traps of the shale surfaces. This could be caused by uncertainties in seismic imaging.

\subsection{Inference setup and posterior parameter estimates}\label{seq:sleipner_parameter_estimates}

Each barrier is assigned one effective entry pressure, giving nine inferred parameters in total. Localized feeders are not represented in this study. However, they could be included by setting the local capillary entry pressure to zero in traps corresponding to the feeder locations identified by \citet{martinez_unraveling_2026}. The per-layer \COtwo{} densities are set according to \citet[Appendix A]{callioli_santi_invasion_2025}, the brine density to $1000\,\mathrm{kg}/\mathrm{m}^3$, the sand porosity to $36\%$, and the irreducible water saturation to $30\%$ \citep{jackson_characterizing_2018}. The observation threshold is $h_{\mathrm{det}}=0$. We select log-normal priors
\[
  \log_{10} P_{\mathrm{th},i}\sim\mathcal{N}\left(1.824,0.326^2\right),
  \qquad
  \log_{10}\lambda\sim\mathcal{N}\left(-11,1^2\right),
\]
where $P_{\mathrm{th},i}$ is expressed in $\mathrm{kPa}$ and $\lambda$ in $\mathrm{m^2\,Pa^{-1}\,s^{-1}}$. The \texttt{PyMC} ABC-SMC sampler uses \texttt{threshold}$=0.75$ and \texttt{correlation\_threshold}$=0.01$. The discrepancy scales for $s_{1, i, k}$ is set to $0.3$ times the corresponding observed area fraction, while to $0.8$ for $s_{2, i, k}$ and $s_{3, i, k}$.  The high value for the lateral summaries reflects the lateral mismatch, as we cannot expect the model to reproduce the observed lateral distribution of the plume. Five independent chains contribute $1000$ final particles each, giving $5000$ posterior samples in total.

\begin{figure}[H]
  \centering
  \includegraphics[width=0.7\linewidth]{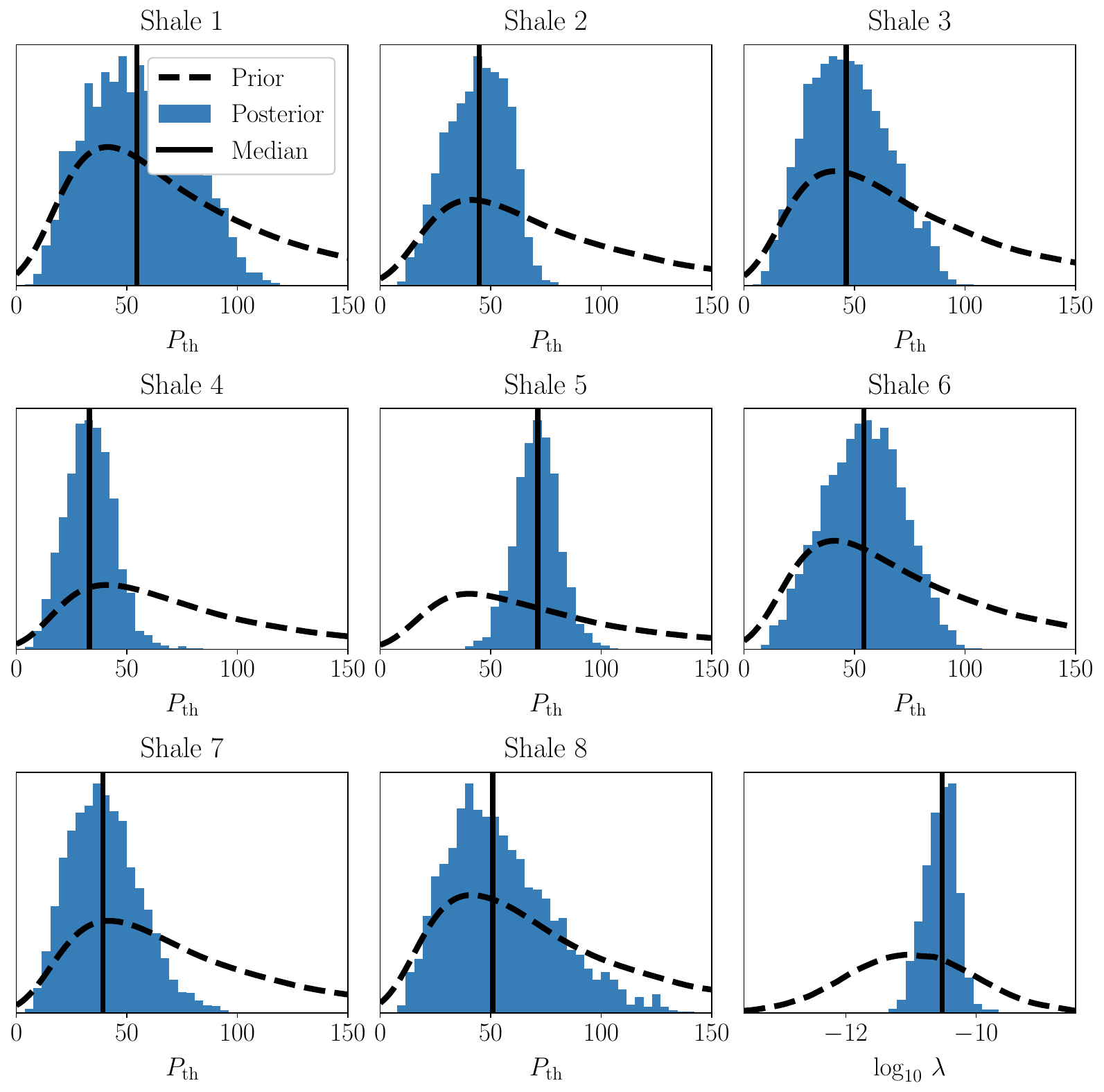}
  \caption{ABC posterior distributions estimated from the Sleipner data using a finite-rate model. Results are shown for the eight shale entry pressures and the shared mobility. Solid vertical lines indicate posterior medians, while dashed lines indicate the log-normal priors.}
  \label{fig:SleipnerPosteriorHistograms}
\end{figure}

\Cref{fig:SleipnerPosteriorHistograms} compares the marginal posterior distributions with the prior distributions. The posterior for $P_{\mathrm{th},5}$ shifts towards higher pressures and contracts, while $P_{\mathrm{th},8}$ remains broad. The mobility posterior also contracts. This suggests the model is able to identify which aspects of the reduced model are identified and which remain unresolved by the available monitoring data. However, note that narrowing distributions do not imply independent parameter identification. We see in Appendix \ref{seq:sleipner_posterior_dependence} that the posterior correlations of $P_{\mathrm{th},5}$ and $P_{\mathrm{th},8}$ with $\log_{10}\lambda$ are $0.72$ and $0.66$, respectively.

\subsection{Posterior reconstruction and model adequacy}

\begin{figure}[H]
  \centering

  \begin{subfigure}[t]{0.48\linewidth}
    \centering
    \includegraphics[width=\linewidth]{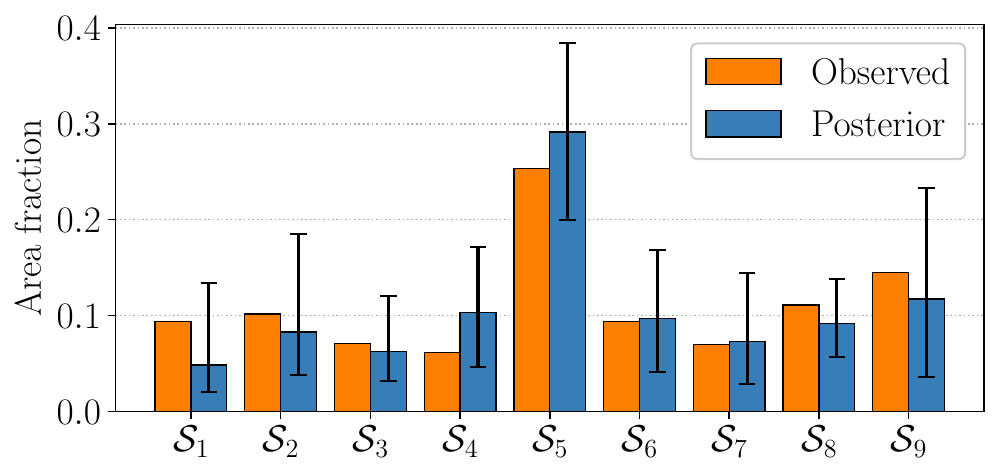}
    \caption{2010}
    \label{fig:SleipnerAreaFractions2010}
  \end{subfigure}
  \hfill
  \begin{subfigure}[t]{0.48\linewidth}
    \centering
    \includegraphics[width=\linewidth]{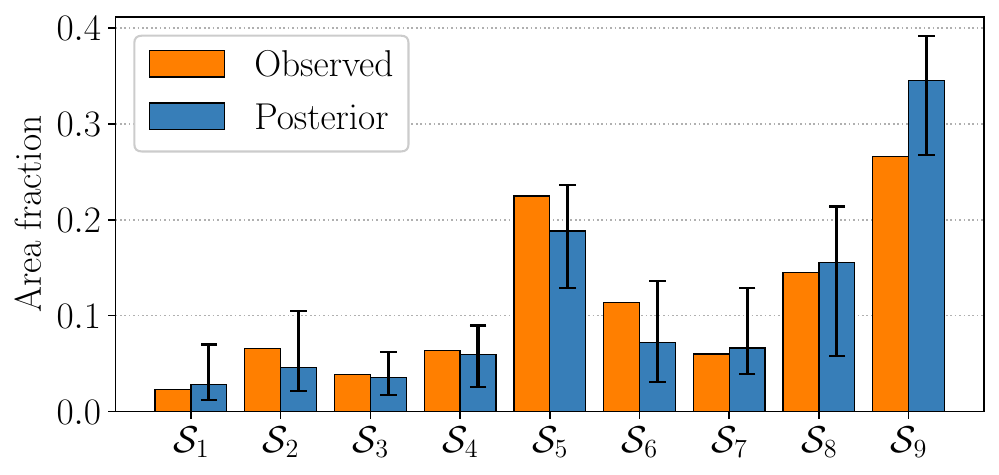}
    \caption{2023}
    \label{fig:SleipnerAreaFractions2023}
  \end{subfigure}

  \caption{Observed and posterior fractions of total detected plume area in each sand unit of Sleipner. The fit uses a finite-rate model. Error bars show $90\%$ posterior reconstruction intervals.}
  \label{fig:SleipnerAreaFractions}
\end{figure}

We pick $500$ of the $5000$ posterior samples, and use them in forward simulations. \Cref{fig:SleipnerAreaFractions} compares the observed footprint-area fractions with their posterior reconstruction distributions for survey times $2010$ and $2023$. The displays indicate how well the model captures the vertical distributions of detected area. Most observations lie within the $90\%$ reconstruction intervals. The only exception is $\mathcal{S}_9$ in $2023$, where the observed fraction lies just below the interval. Importantly, the main temporal shift is also reproduced. Storage unit $\mathcal{S}_5$ has the largest observed footprint fraction in $2010$, while a larger share of the detected area occupies $\mathcal{S}_9$ by $2023$. Appendix \ref{seq:sleipner_observation_times} compares how the forecast layer-wise mass fractions change when using only the $2010$ observation or both $2010$ and $2023$.

\begin{figure}[H]
  \centering

  \begin{subfigure}[t]{0.48\linewidth}
    \centering
    \includegraphics[width=\linewidth]{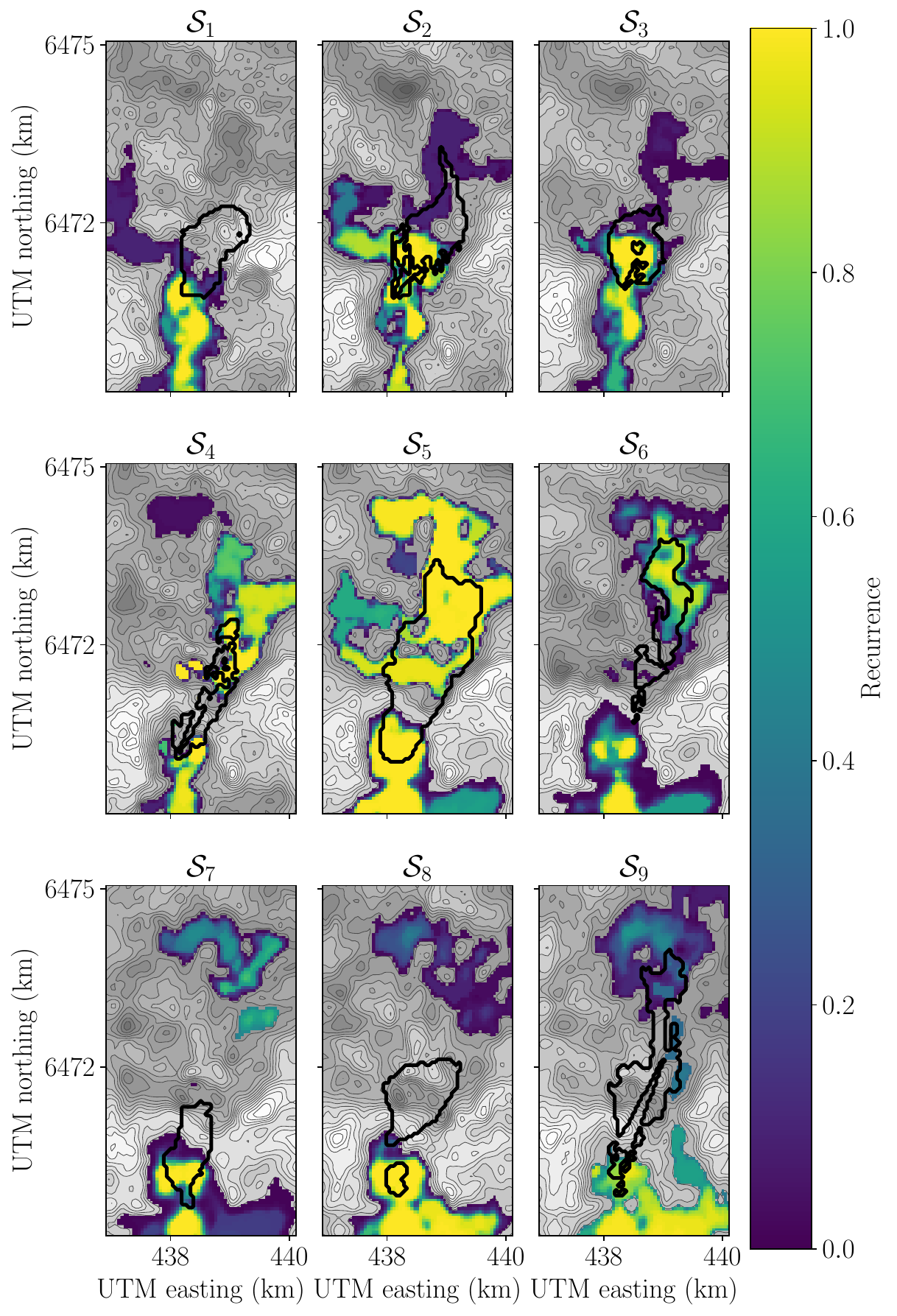}
    \caption{2010}
    \label{fig:SleipnerMeanPredictions2010}
  \end{subfigure}
  \hfill
  \begin{subfigure}[t]{0.48\linewidth}
    \centering
    \includegraphics[width=\linewidth]{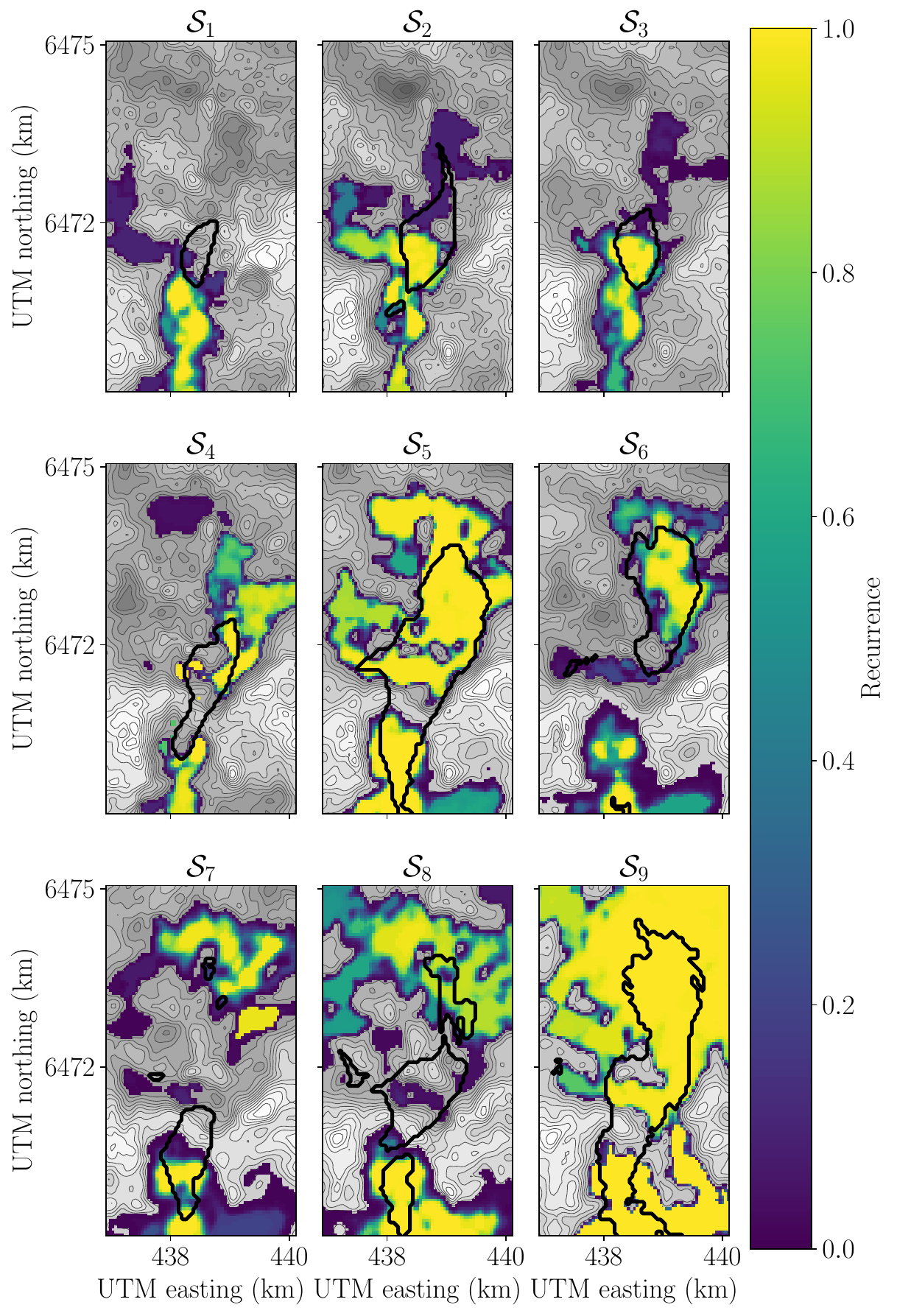}
    \caption{2023}
    \label{fig:SleipnerMeanPredictions2023}
  \end{subfigure}

  \caption{Posterior recurrence of the detected plume at the two survey times used for the Sleipner fit. Units $\mathcal{S}_1, \ldots, \mathcal{S}_9$ are ordered from deepest to shallowest. Colour shows the fraction of posterior simulations containing detected \COtwo{} in each cell, and black lines show the observed outlines.}
  \label{fig:SleipnerMeanPredictions}
\end{figure}

Using the observation rule from \Cref{seq:seismic_detection}, each posterior simulation is converted to a binary plume-detection mask. Cell-wise recurrence is the fraction of these simulations in which \COtwo{} is detected in a given cell, with values near one indicating consistent plume presence across the posterior ensemble. \Cref{fig:SleipnerMeanPredictions} shows that the simulations generally place \COtwo{} in all nine sand units by 2010, reproduce the large accumulation in $\mathcal{S}_5$, and concentrate later footprint growth in the upper units. Mismatch between the assumed topography and observed plume geometry leads to weaker lateral agreement. In several middle and upper units, the simulated plume lies south of the observed outline or splits between competing structural traps. This is consistent with \citet{callioli_santi_invasion_2025}, who also found southward migration in continuous-shale models.

\subsection{Comparison with quasi-static transfer}\label{seq:sleipner_quasi}

\begin{figure}[H]
  \centering

  \begin{subfigure}[t]{0.48\linewidth}
    \centering
    \includegraphics[width=\linewidth]{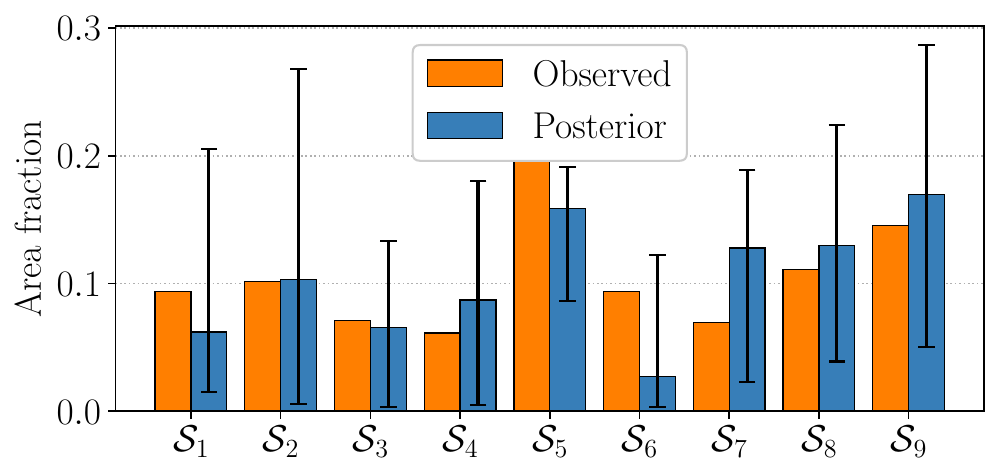}
    \caption{2010}
    \label{fig:SleipnerQuasiAreaFractions2010}
  \end{subfigure}
  \hfill
  \begin{subfigure}[t]{0.48\linewidth}
    \centering
    \includegraphics[width=\linewidth]{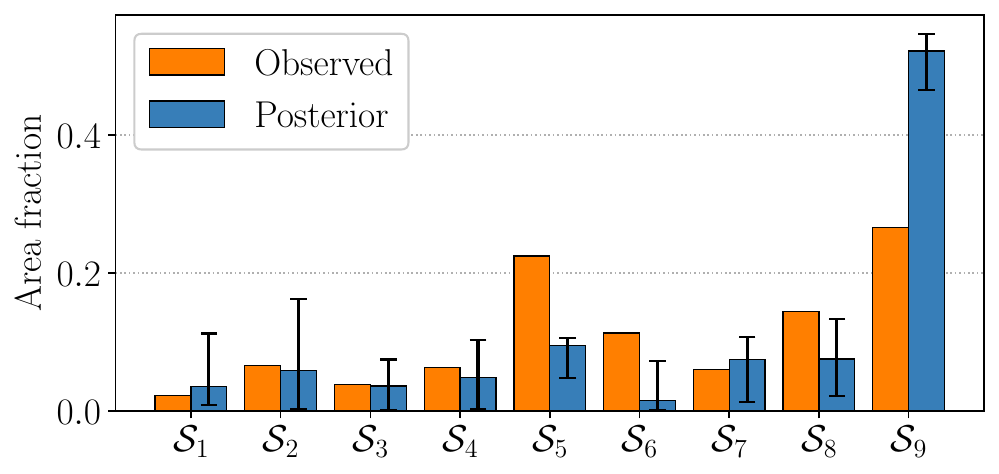}
    \caption{2023}
    \label{fig:SleipnerQuasiAreaFractions2023}
  \end{subfigure}

  \caption{Observed and posterior fractions of total detected plume area in each sand unit of Sleipner. The fit uses a quasi-static model. Error bars show $90\%$ posterior reconstruction intervals.}
  \label{fig:SleipnerQuasiAreaFractions}
\end{figure}

The role of finite-rate transfer is assessed by refitting the site with a quasi-static IP model. We use the same priors and ABC-parameters are earlier, but omit the rate-limitation. \Cref{fig:SleipnerQuasiAreaFractions} shows that this model cannot reproduce the layer-wise distribution of detected \COtwo{} area at both survey times. Only the top layer evolves between $2010$ and $2023$, which leads to overestimation of the detected area in that layer. The quasi-static fit also compensates for its inability to retain \COtwo{} beneath a breached barrier by inferring systematically higher effective entry pressures. Posteriors can be found in Appendix \ref{seq:sleipner_quasi_static}, and medians range from $81$ to $118\,\mathrm{kPa}$, compared with $33$-$71\,\mathrm{kPa}$ for the finite-rate model. Note that these higher values represent compensation within the reduced model rather than direct estimates of physical shale entry pressures.

\section{Application to a synthetic multilayer model}\label{seq:synthetic}

\subsection{Experimental design}

The Sleipner application tests whether the framework can be conditioned on an observed multilayer plume, but the unknown shale properties prevent direct assessment of parameter accuracy. Synthetic experiments provide a complementary setting in which the true parameters and future plume evolution are known by construction, allowing direct evaluation of parameter recovery, out-of-sample forecasting, and sensitivity to monitoring duration and model assumptions. These tests remain conditional on the assumed model and simulator and therefore do not assess robustness to broader model discrepancy.

The synthetic reservoir contains four sand units $\mathcal{S}_{1}, \ldots, \mathcal{S}_{4}$ separated by shale barriers. It has a $5\,\mathrm{km}\times5\,\mathrm{km}$ domain is discretised on a $51\times51$ lateral grid. \COtwo{} is injected at a constant rate into the centre of the deepest unit. Each sand unit is $50\,\mathrm{m}$ thick and each shale is $6\,\mathrm{m}$ thick, with the top of the stack at $800\,\mathrm{m}$ depth. The four structural surfaces are independent realisations of a Gaussian random field (GRF) \citep{banerjee2025hierarchical} using a Mat\'ern covariance with smoothness $\nu=3/2$, correlation length $1\,\mathrm{km}$ and standard deviation $10\,\mathrm{m}$. Sand porosity is $36\%$, the irreducible water saturation is $30\%$, and the \COtwo{} and brine densities are $700$ and $1000\,\mathrm{kg\,m^{-3}}$, respectively. The true parameters as selected as $(P_{\mathrm{th}, 1},P_{\mathrm{th}, 2},P_{\mathrm{th}, 3})=(44,52,66)\,\mathrm{kPa}$ and $\lambda=10^{-10}\,\mathrm{m^2\,Pa^{-1}\,s^{-1}}$.

\begin{figure}[H]
  \centering
  \includegraphics[width=0.85\linewidth]{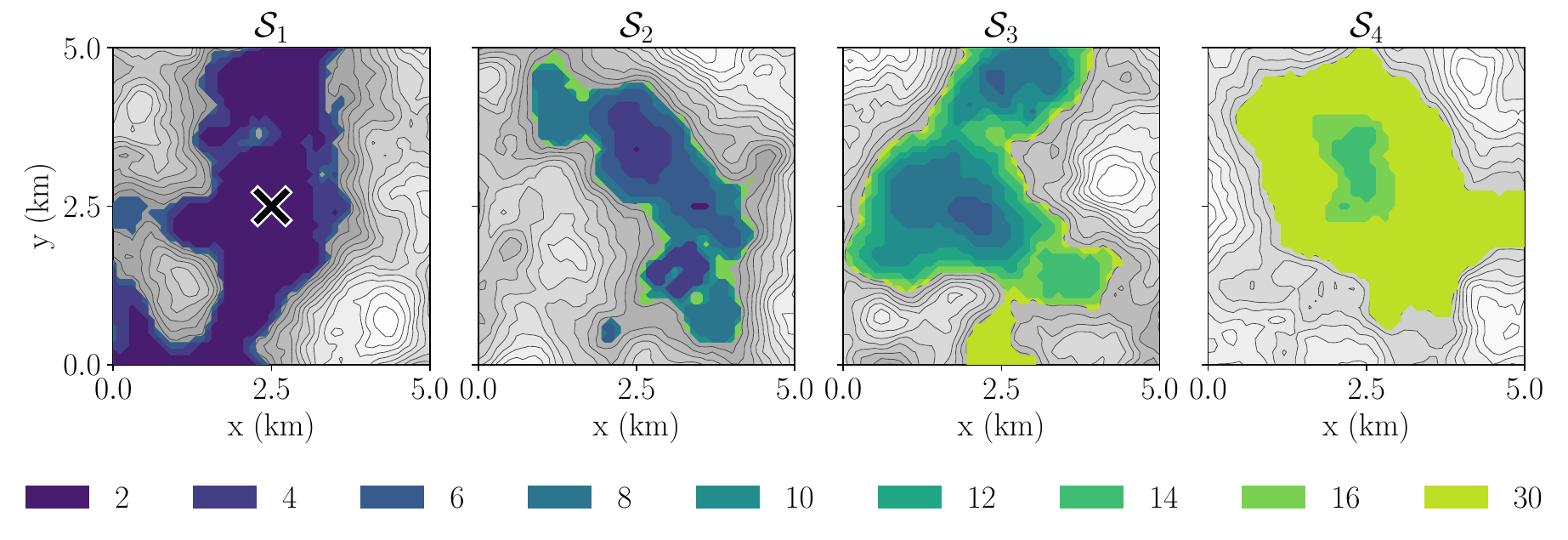}
  \caption{True plume evolution in the GRF reservoir. Units $\mathcal{S}_1,\ldots,\mathcal{S}_4$ are ordered from deepest to shallowest. Coloured plume extents are shown from year $2$ to year $16$, together with the forecast target at year $30$.}
  \label{fig:synth_truth_birds_eye}
\end{figure}

We evaluate effect of monitoring duration by considering two different observation scenarios. Scenario $1$ surveys the reservoir every two years from year $2$ through year $16$. Scenario $2$ uses the same truth, but includes only the first five surveys and ends in year $10$. \Cref{fig:synth_truth_birds_eye} shows plume evolution through year $16$ and a later forecast target at year $30$. All three barriers are crossed during scenario $1$, with \COtwo{} first entering $\mathcal{S}_4$ in year $13$. In scenario $2$, the shallowest barrier therefore remains unbreached during the observation period.

Unless stated otherwise, the detection threshold is set to $h_{\mathrm{det}}=0$. The entry-pressures are given priors $\log_{10} P_{\mathrm{th},i}\sim\mathcal{N}\left(1.732,0.304^2\right)$, centred at the mean $54\,\mathrm{kPa}$ of the three true values. The mobility prior is $\log_{10}\lambda\sim\mathcal{N}\left(-9.523,1.0^2 \right)$, with a median three times the true value. Recovery therefore cannot be attributed to priors centred on the truth. We set discrepancy scales to $0.1$ times the observed area fractions and $0.3$ times the observed spatial moments. The ABC-SMC sampler in \texttt{PyMC} uses the same \texttt{threshold} and \texttt{correlation\_threshold} as in \Cref{seq:sleipner}. Five independent chains of $1000$ particles yield $5000$ final particles.

\subsection{Parameter recovery}\label{seq:synthetic_estimation}

\begin{figure}[H]
  \centering
  \begin{subfigure}[t]{\linewidth}
    \centering
    \includegraphics[width=\linewidth]{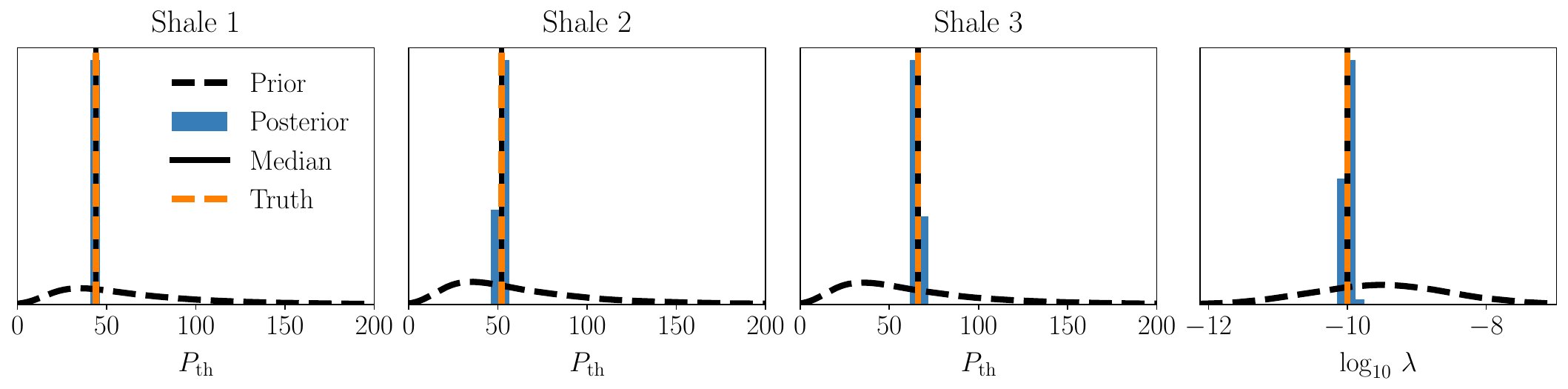}
    \caption{Scenario $1$. GRF reservoir model observed to year $16$.}
    \label{fig:synth_grf_recovery}
  \end{subfigure}
  \vfill
  \begin{subfigure}[t]{\linewidth}
    \centering
    \includegraphics[width=\linewidth]{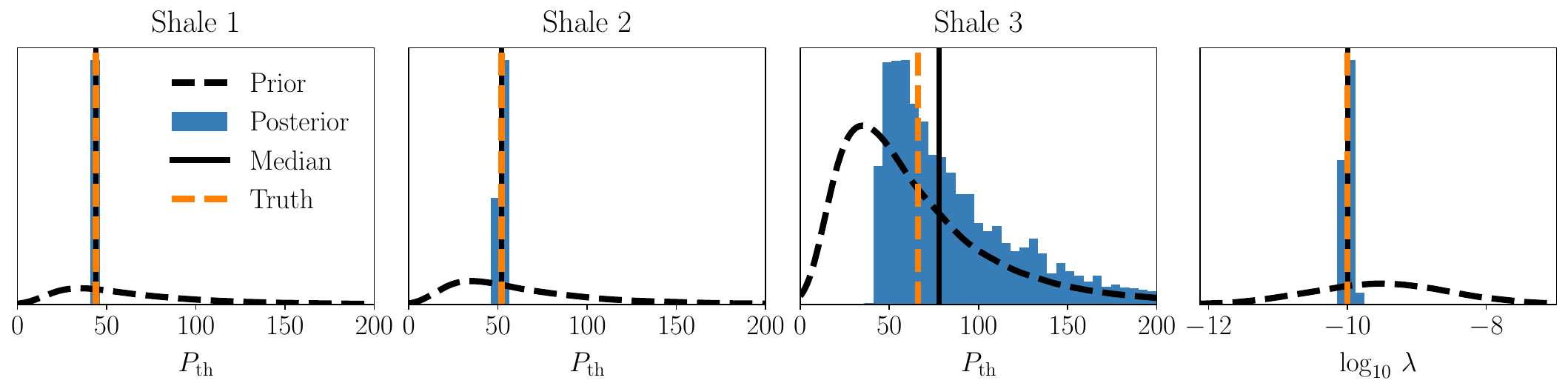}
    \caption{Scenario $2$. GRF reservoir model observed to year $10$.}
    \label{fig:synth_grf_short_recovery}
  \end{subfigure}

  \caption{Marginal priors and ABC posteriors for the two GRF scenarios. Dashed black curves show the priors, blue histograms show the posteriors, solid black lines show the posterior medians, and dashed orange lines show the true values.}
  \label{fig:synth_recovery}
\end{figure}

Fitting the model, \Cref{fig:synth_recovery} compares the marginal priors and ABC posteriors with the true parameters. In scenario~$1$, all three barriers are crossed during monitoring, so all four posteriors contract strongly around the truth. The entry-pressure $90\%$ intervals have widths of $1.3$--$4.1\,\mathrm{kPa}$. Scenario $2$ gives similar estimates of $P_{\mathrm{th}, 1}$, $P_{\mathrm{th}, 2}$ and $\lambda$, but the posterior for $P_{\mathrm{th},3}$ is broad and right-skewed. Its median lies $18\%$ above the truth and its $90\%$ interval spans $46$-$200\,\mathrm{kPa}$.  This asymmetry reflects that $P_{\mathrm{th},3}$ is informed primarily through a lower bound, since values low enough to trigger an earlier crossing are disfavoured. The results therefore suggest that information gain is event-dependent rather than only a function of the number of surveys. Observations before a barrier is crossed can rule out thresholds that would imply an earlier crossing, but do not strongly constrain larger values.

\subsection{Forecasting plume migration}\label{seq:synthetic_forecast}

\begin{figure}[H]
  \centering
  \begin{subfigure}[t]{\linewidth}
    \centering
    \includegraphics[width=\linewidth]{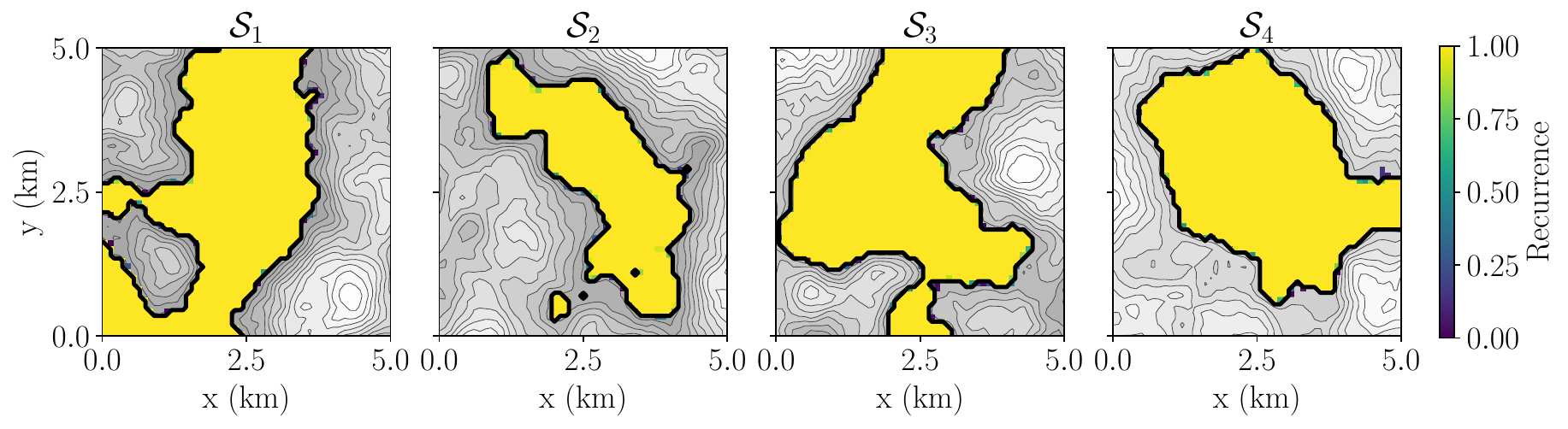}
    \caption{Scenario $1$. GRF observed to year $16$.}
    \label{fig:synth_grf_forecast_probabilities}
  \end{subfigure}
  \vfill
  \begin{subfigure}[t]{\linewidth}
    \centering
    \includegraphics[width=\linewidth]{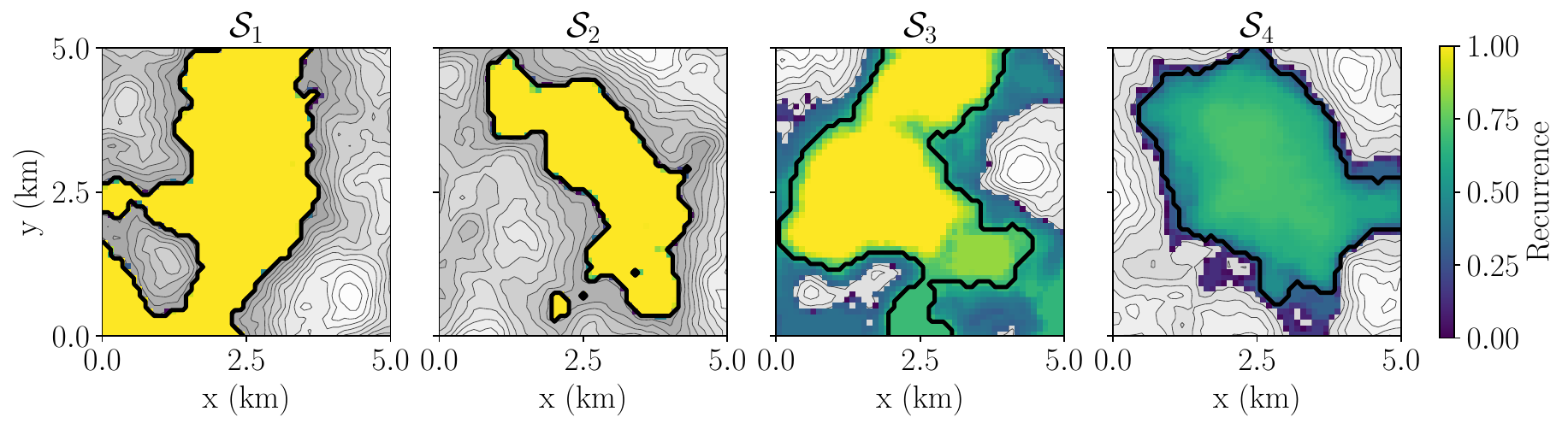}
    \caption{Scenario $2$. GRF observed to year $10$.}
    \label{fig:synth_grf_short_forecast_probabilities}
  \end{subfigure}

  \caption{Posterior plume recurrence at year $30$, estimated from $500$ posterior simulations. Black lines show the true plume outlines. Cells with zero recurrence are not coloured.}
  \label{fig:synth_forecast_probabilities}
\end{figure}

For each scenario, $500$ randomly selected posterior samples are propagated from the final survey and forwards in time. \Cref{fig:synth_forecast_probabilities} shows the resulting recurrence maps for year $30$. Scenario~$1$ has high recurrence within the true outlines and sharp transitions at their boundaries in all units. Scenario $2$ consistently reproduces the two deepest units, whereas $\mathcal{S}_3$ and $\mathcal{S}_{4}$ have broad transition zones.

\begin{figure}[H]
  \centering
  \begin{subfigure}[t]{\linewidth}
    \centering
    \includegraphics[width=\linewidth]{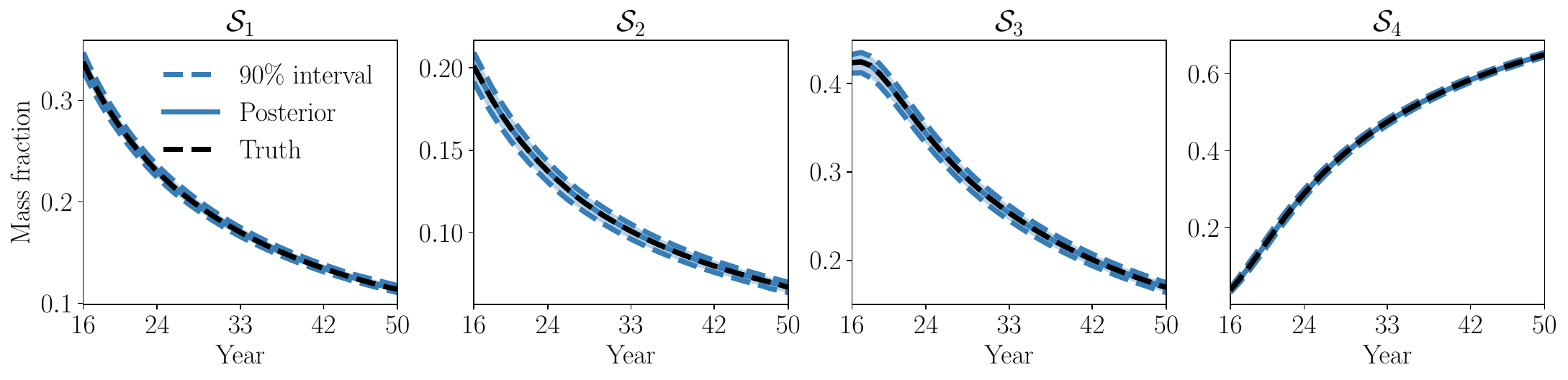}
    \caption{Scenario $1$. GRF observed to year $16$.}
    \label{fig:synth_grf_forecast_mass_timeseries}
  \end{subfigure}
  \vfill
  \begin{subfigure}[t]{\linewidth}
    \centering
    \includegraphics[width=\linewidth]{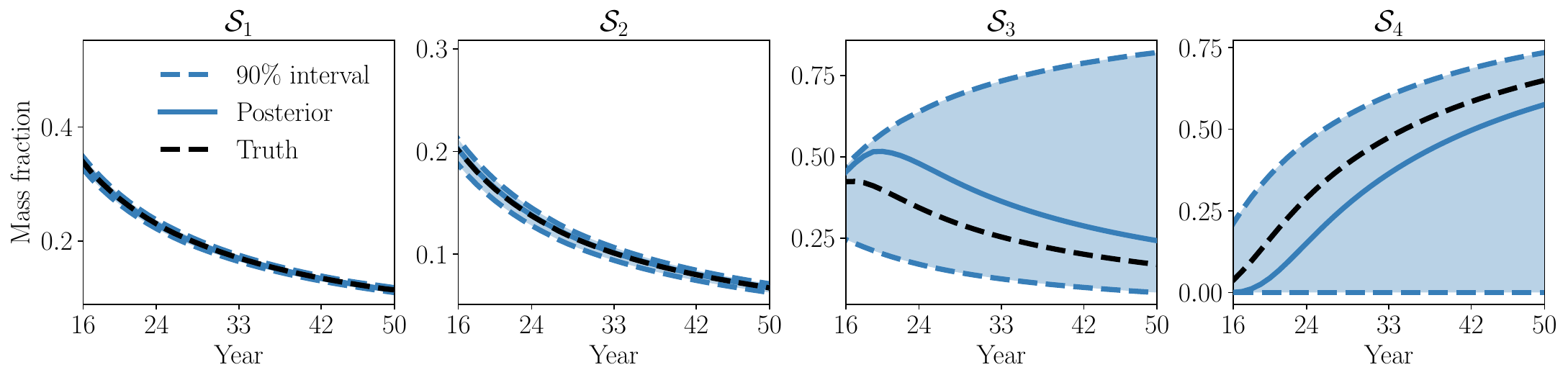}
    \caption{Scenario $2$. GRF observed to year $10$.}
    \label{fig:synth_grf_short_forecast_mass_timeseries}
  \end{subfigure}

  \caption{Forecast fractions of total stored \COtwo{} mass in each sand unit. Solid blue lines show posterior medians, blue bands show $90\%$ posterior forecast intervals, and dashed black lines show the truth.}
  \label{fig:synth_forecast_mass_timeseries}
\end{figure}

\Cref{fig:synth_forecast_mass_timeseries} shows the fraction of the total stored mass in each sand unit, forecast annually from the end of the observation period to year $50$. The truth lies within the $90\%$ posterior forecast interval at every plotted time in both scenarios. At year $30$, scenario $1$ has an average interval width of $0.014$ and a root-mean-square error of $0.0004$ for the posterior median across the four units. The corresponding values in scenario $2$ are $0.290$ and $0.0849$. Thus, ending observations six years earlier increases these metrics by factors of about $20$ and $200$, respectively. The additional uncertainty is concentrated in $\mathcal{S}_3$ and $\mathcal{S}_4$ because the weak constraint on $P_3$ leaves the timing of the third barrier breach uncertain. See that the uncertainty intervals in $\mathcal{S}_{4}$ include the scenario where no \COtwo{} reaches the shallowest unit, which is consistent with the low recurrence from \Cref{fig:synth_grf_short_forecast_probabilities}. 

\subsection{Accounting for the finite-rate transfer}\label{seq:synthetic_transfer}

The experiments above use the finite-rate transfer law of \Cref{eq:seal_flux} both to generate and fit the synthetic data. We now perform a controlled model-discrepancy test where we fit both the finite-rate and quasi-static models to a truth from a finite-rate scenario. The same experiment is also repeated with a quasi-static truth, and the results are included in Appendix \ref{seq:synthetic_transfer_additional_results}. Settings for the reservoir, true parameters and ABC inference are mostly unchanged. but the finite-rate mobility prior is widened to $\log_{10}\lambda\sim\mathcal{N}\left(-9.0,2.0^2\right)$ to include a quasi-static limit.

\begin{figure}[H]
  \centering
  \begin{subfigure}[t]{\linewidth}
    \centering
    \includegraphics[width=\linewidth]{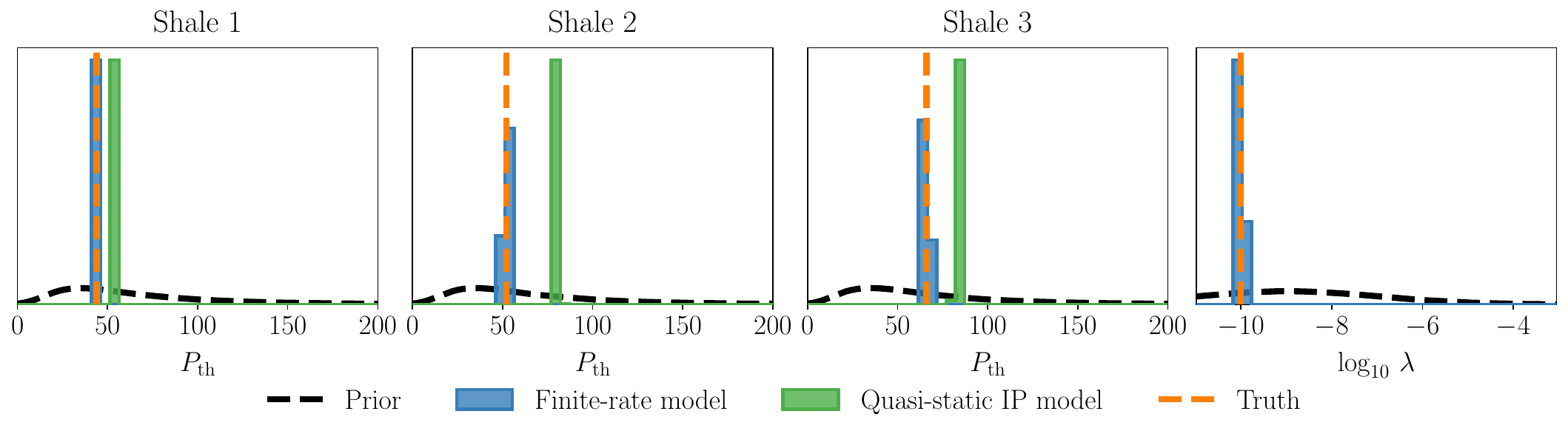}
    \caption{Posterior distributions}
    \label{fig:synth_transfer_rate_posteriors}
  \end{subfigure}
  \vfill
  \begin{subfigure}[t]{\linewidth}
    \centering
    \includegraphics[width=\linewidth]{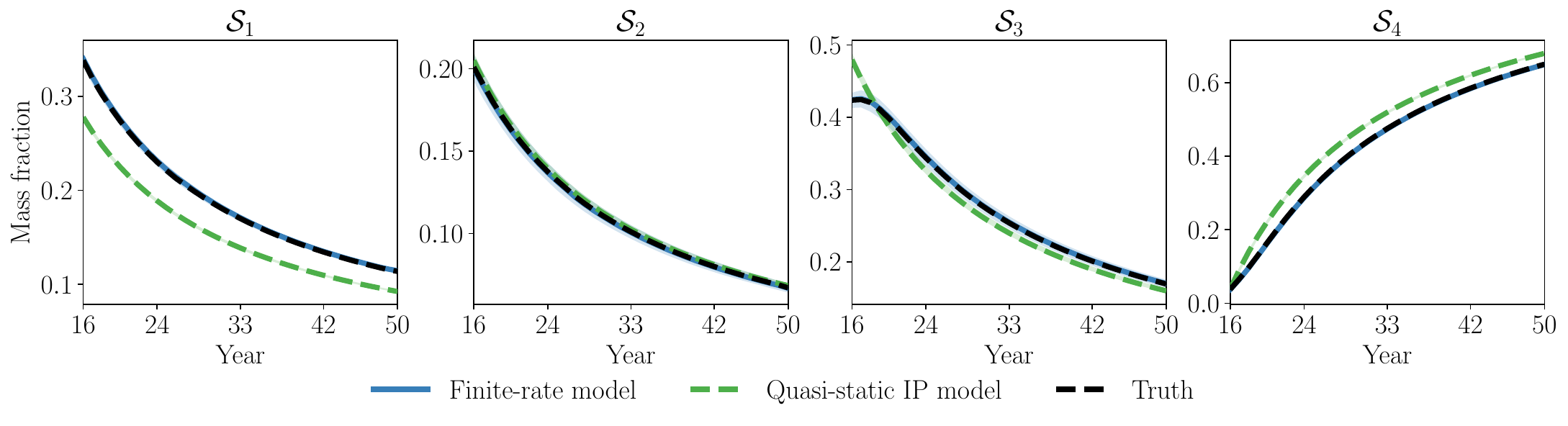}
    \caption{Forecast mass fractions}
  \end{subfigure}

  \caption{Finite-rate and quasi-static fits to the finite-rate synthetic truth. In (a), dashed black curves show the shared entry-pressure priors and dashed orange lines show the true parameters. In (b), the models are compared with the true layer-wise mass fractions.}
  \label{fig:synth_transfer_rate}
\end{figure}

\Cref{fig:synth_transfer_rate} compares posterior distributions and forecasts under the finite-rate truth. The finite-rate model recovers the entry pressures and reproduces the true layer-wise mass fractions. The quasi-static model cannot retain \COtwo{} beneath a breached barrier and compensates by inferring higher entry pressures that delay upward migration. Yet, even with this compensation, the forecast underestimates mass retention in the lowest unit.

\subsection{Sensitivity to the plume-detection threshold}\label{seq:plume_detection}

The preceding experiments assume $h_{\mathrm{det}}=0$, so the observed footprint equals the full simulated plume footprint. To assess sensitivity to uncertain detectability, observations are generated with a true threshold of $3\,\mathrm{m}$ and fitted using assumed thresholds of $0$, $3$ and $8\,\mathrm{m}$. Appendix \ref{seq:plume_detection_extents} illustrates the geometric effect of these thresholds, where we see that increasing the threshold progressively removes thin plume margins and small accumulations. 

\begin{figure}[H]
  \centering
  \includegraphics[width=\linewidth]{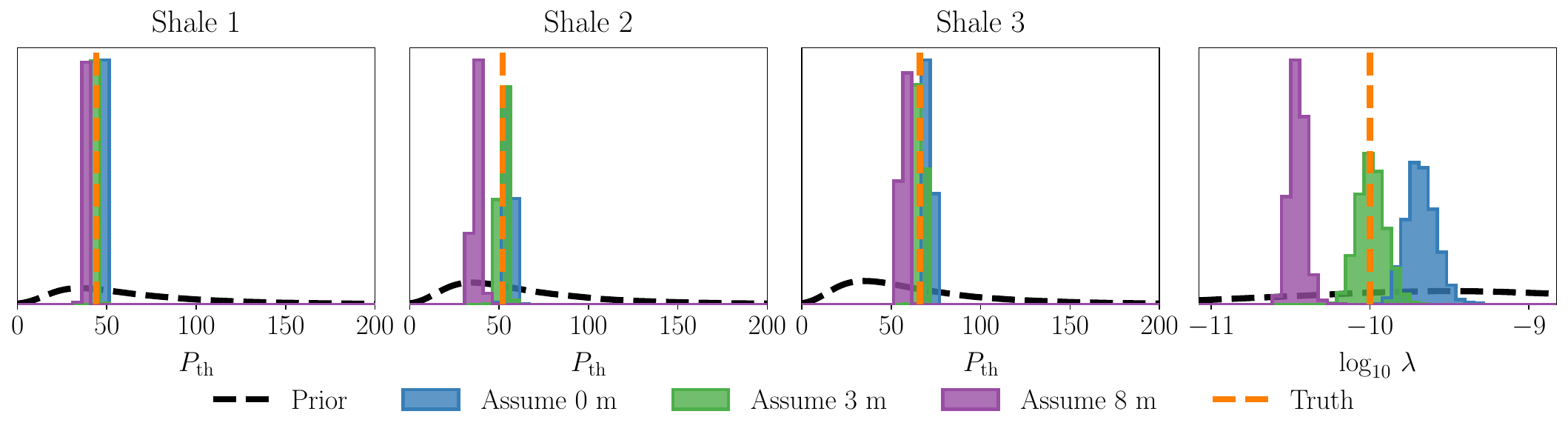}
  \caption{Posteriors obtained with assumed detection thresholds of $0$, $3$ and $8\,\mathrm{m}$ when the true threshold is $3\,\mathrm{m}$. Dashed black curves show the priors and dashed orange lines show the true parameters.}
  \label{fig:synth_detection_posterior}
\end{figure}

\Cref{fig:synth_detection_posterior} shows that misspecifying $h_{\mathrm{det}}$ biases the inferred parameters. The entry-pressure posteriors are similar for assumed thresholds of $0$ and $3,\mathrm{m}$, whereas the $8,\mathrm{m}$ scenario shifts towards lower pressures. Relative to the correctly specified $h_{\mathrm{det}}$, the mobility posterior shifts towards lower values for $h_{\mathrm{det}}=0$ and higher for $h_{\mathrm{det}}=8,\mathrm{m}$. This indicates that misspecified detectability can be absorbed by both effective barrier strength and mobility. Posterior contraction should thus be interpreted together with the assumed observation model, since detection errors can appear as changes in the effective migration parameters.

\section{Closing remarks}\label{seq:closing_remarks}

We have presented a Bayesian framework for estimating effective parameters that control vertical \COtwo{} migration in layered reservoirs. The framework targets buoyancy- and capillary-dominated settings where full-physics inference is too costly or an interpretable reduced model is preferred. It develops a graph-based finite-rate IP simulator, and combines it with ABC-SMC for parameter estimation from time-lapse seismic plume outlines. For fixed reservoir geometry, the graph is constructed once and reused across parameter evaluations. On the reposted benchmark, a finite-rate forward simulation had a median runtime of $64.2\,\mathrm{ms}$, making Monte Carlo-based inference computationally tractable.

The Sleipner experiment tests the framework under real data and model limitations, rather than against known parameter values. We find the posterior simulations to reproduce the broad vertical distribution of the detected plume across nine sand units and the shift in footprint area towards the upper units between $2010$ and $2023$. The surveys narrow several marginal parameter distributions, but posterior dependence indicates that effective barrier strength and mobility are not independently identified. Lateral agreement between observed and simulated outlines is weaker, reflecting both the topography-trap mismatch and the omission of localized feeders.

Synthetic experiments provide complementary tests of parameter recovery and out-of-sample forecasting, and show how uncertainty can be reduced as additional monitoring data are acquired. When monitoring continues until every barrier has been crossed, the posteriors concentrate around the true parameters and forecasts closely reproduce future plume states. Ending the observation period before the final crossing instead leaves the corresponding threshold pressure weakly constrained and increases forecast interval widths. Fitting quasi-static transfer to finite-rate observations inflates the inferred threshold pressures by more than $50\%$ for the most affected barrier, whereas the finite-rate model recovers the truth and approaches the quasi-static limit when transfer is instantaneous. Misspecifying the plume-detection threshold biases both the inferred entry pressures and mobility, which motivates explicit treatment of seismic detectability in future applications.

Together, these results suggest a practical role for the framework during active injection. As new monitoring surveys become available, the inference can be updated to revise the effective migration parameters and corresponding plume forecasts. The value of each survey depends on the migration events it captures, with breakthrough and post-breach redistribution providing particularly strong constraints. Remaining posterior uncertainty then highlights aspects of the migration behaviour that are still unresolved by the available observations.

Future work should test robustness to uncertain shale topography and unknown localized shale breaks, and could infer additional observation-model parameters such as $h_{\mathrm{det}}$. We only used plume outlines data in this study, and more informative seismic data and summary statistics may improve inference. Benchmarking the simulator against laboratory measurements and higher-fidelity flow simulations would clarify its fidelity to the underlying physics. This would also inform which additional physical processes should be included in the model. Furthermore, applying the framework to other storage sites can show how it performs under different geological settings and monitoring conditions. Lastly, as the fast simulator that was shown to match real-world Sleipner results rather well, we also foresee opportunities for using it as a surrogate model in for instance optimization of drilling plans, optimal survey design calculations, and other high-level tasks going beyond inversion.

\begin{appendices}
\section{Supporting Sleipner results}\label{seq:sleipner_additional_results}

\subsection{Posterior dependence}\label{seq:sleipner_posterior_dependence}
\begin{figure}[H]
  \centering
  \includegraphics[width=0.55\linewidth]{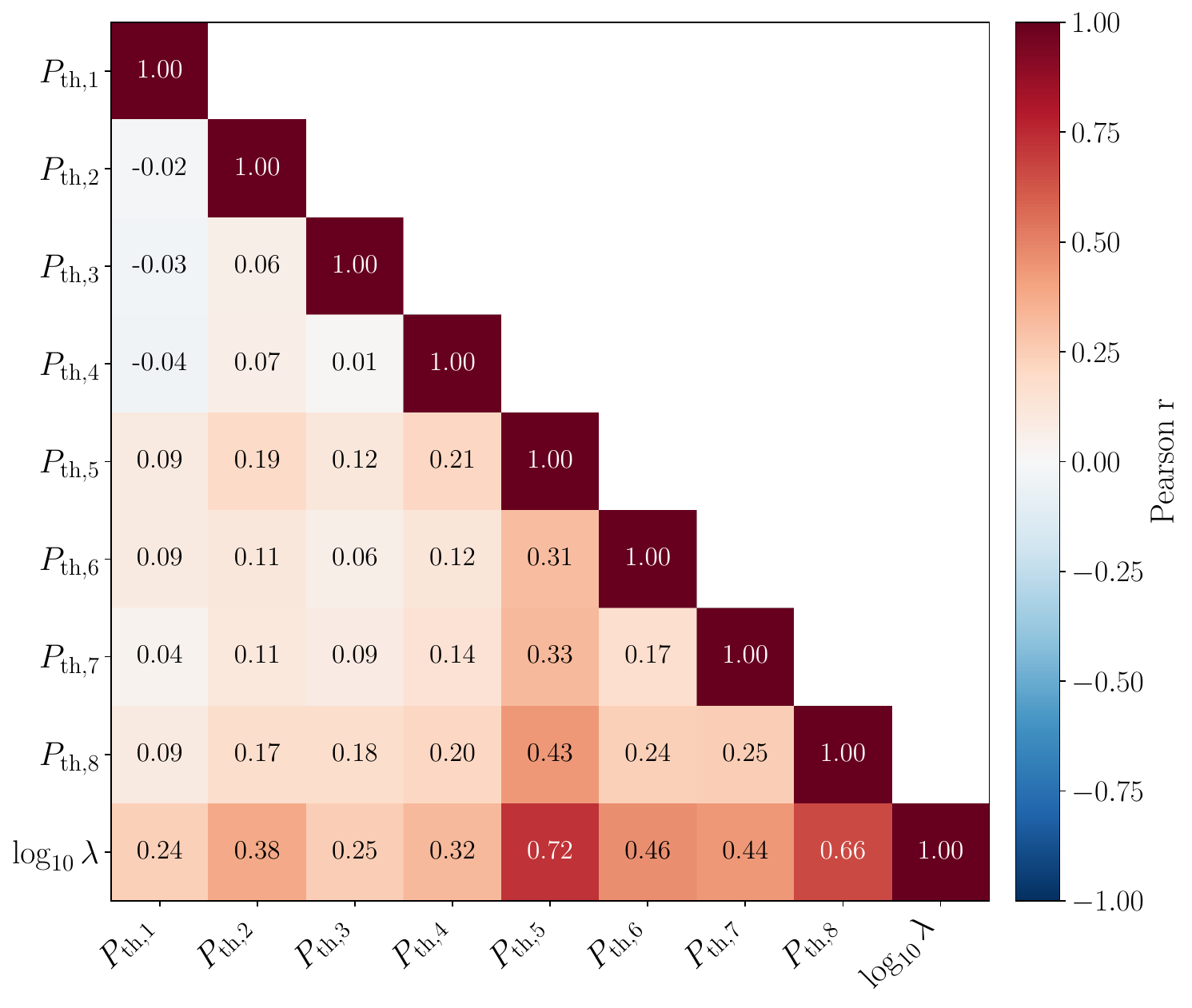}
  \caption{Pearson correlations among the eight entry pressures and $\log_{10}\lambda$ in the Sleipner ABC posterior. Fit using a finite-rate model.}
  
  \label{fig:SleipnerPosteriorCorrelations}
\end{figure}

Faster post-breach transfer can be offset by later breakthrough, producing a trade-off between effective barrier strength and mobility. \Cref{fig:SleipnerPosteriorCorrelations} shows the correlations among the eight entry pressures and $\log_{10}\lambda$ in the estimated Sleipner ABC posterior from \Cref{seq:sleipner_parameter_estimates}. See that the correlations between the capillary entry pressures and $\log_{10} \lambda$ are relatively small for the lower barriers. However, the association is stronger for $P_{\mathrm{th},5}$ and $P_{\mathrm{th},8}$, whose correlations with $\log_{10}\lambda$ are $0.72$ and $0.66$, respectively. These dependencies show that narrow marginal distributions do not imply independent parameter identification.

\subsection{Effect of additional monitoring data}\label{seq:sleipner_observation_times}

\begin{figure}[H]
  \centering
  \includegraphics[width=0.6\linewidth]{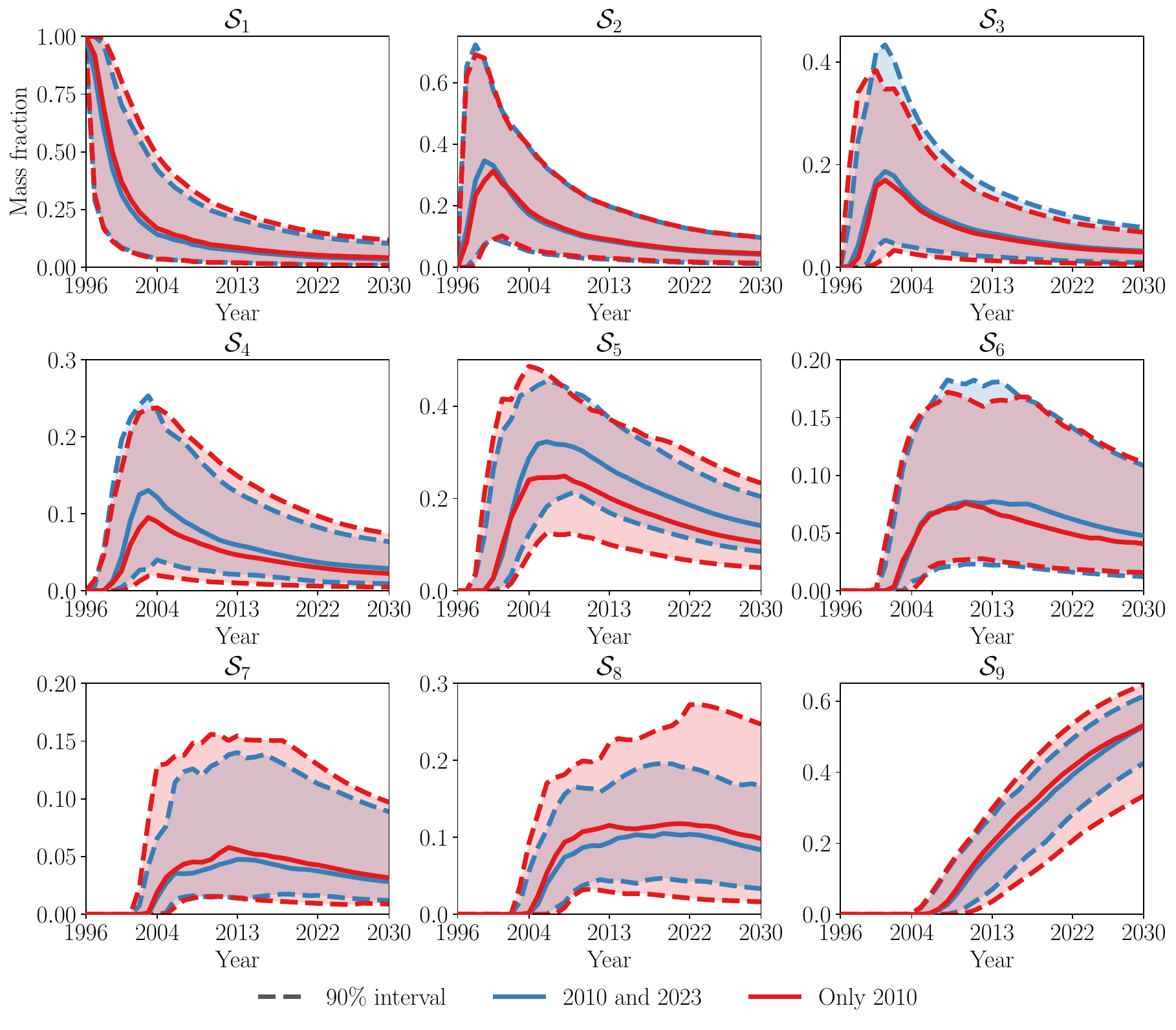}
  \caption{Forecast layer-wise mass fractions for the finite-rate Sleipner model using only the $2010$ observation or both $2010$ and $2023$. Solid lines show posterior medians, and shaded areas show $90\%$ forecast intervals.}
  \label{fig:SleipnerForecastObservationTimes}
\end{figure}

In the main Sleipner experiment, we used both observation times $2010$ and $2023$ for fitting the model. \Cref{fig:SleipnerForecastObservationTimes} compares the forecast layer-wise mass fractions when using only the $2010$ observation or both $2010$ and $2023$. As expected, the two-observation fit produces narrower forecast intervals, but the differences are small for the lower layers. The posterior medians mostly agree, with a notable difference in layer $5$. The comparison therefore shows that the reduction in forecast uncertainty does not uniformly decrease when the monitoring record is extended. Interestingly, both fits forecast that the \COtwo{} will not reach the top layer until after $2004$, while in reality this has already occurred by $1999$. This discrepancy indicates the time of monitoring matters. As the first observation is from $2010$, we have little information about the early migration history, the reduced-order models choose parameters that delay upward migration.

\subsection{Quasi-static parameter estimates}\label{seq:sleipner_quasi_static}

\begin{figure}[H]
  \centering
  \includegraphics[width=0.55\linewidth]{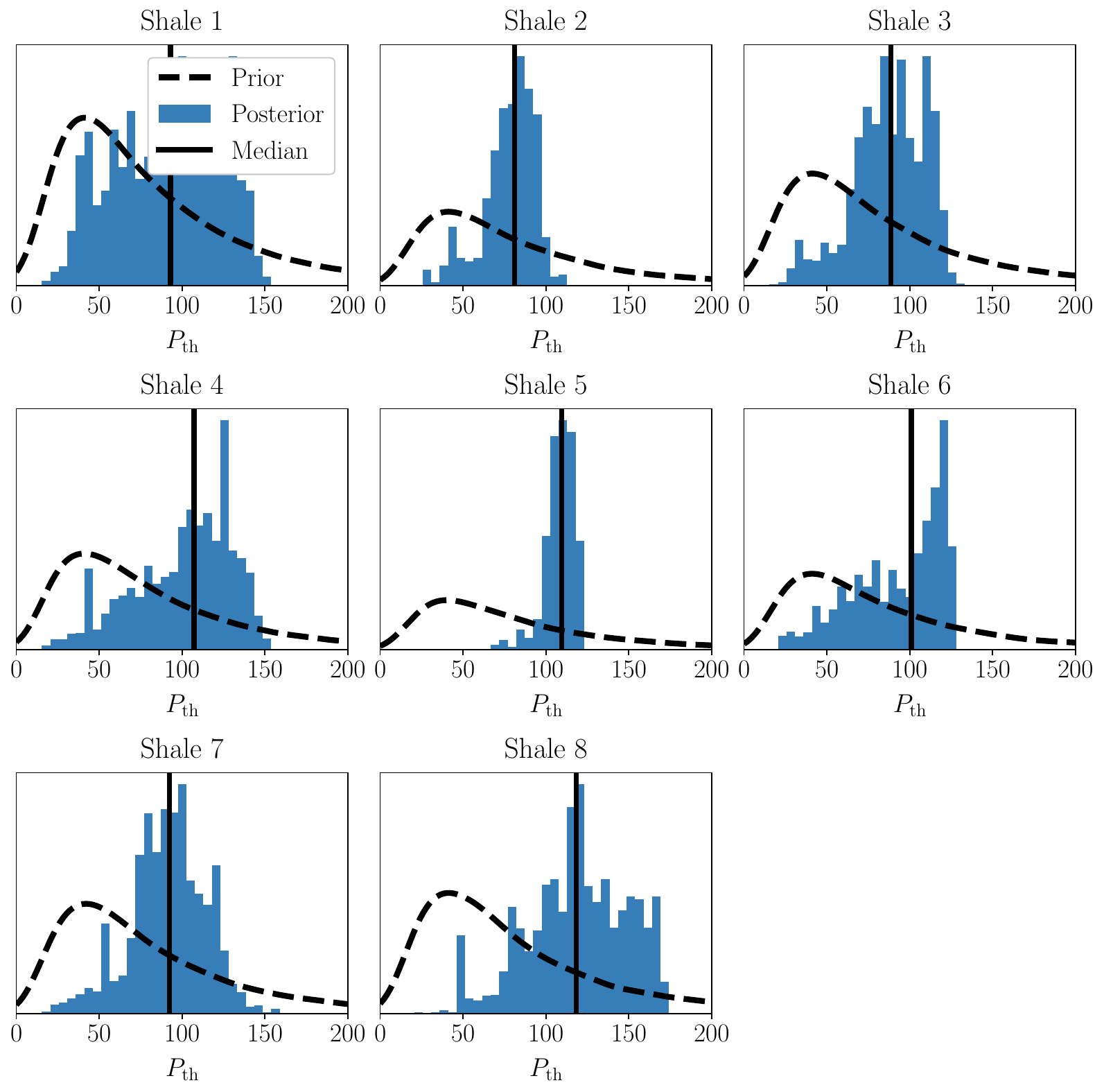}
  \caption{ABC posterior distributions estimated from the Sleipner data using a quasi-static model. Results are shown for the eight shale entry pressures and the shared mobility. Solid vertical lines indicate posterior medians, while dashed lines indicate the log-normal priors.}
  \label{fig:SleipnerQuasiPosteriorHistograms}
\end{figure}

\Cref{seq:sleipner_quasi} compared the finite-rate and quasi-static transfer models for the Sleipner data. We here provide the corresponding posterior distributions for the quasi-static model. See from \Cref{fig:SleipnerQuasiPosteriorHistograms} that the quasi-static model produces higher posterior medians for every barrier than the finite-rate fit. This compensation is done to keep more \COtwo{} in the lower units. Yet, from the forecast we found that this caused a misfit in the time-evolution of the layer-wise mass distribution. Also observe that the posteriors appear less concentrated around the medians. This could be due to the ABC-SMC algorithm struggling to converge, as it is unable to find parameter sets that match the observations.

\section{Supporting synthetic results}\label{seq:synthetic_additional_results}

\subsection{Quasi-static limiting case}\label{seq:synthetic_transfer_additional_results}

The main model-comparison experiment in \Cref{seq:synthetic_transfer} examines the consequences of fitting quasi-static transfer to finite-rate observations. We here provide a complementary test of whether the finite-rate model can fit a quasi-static truth. Recall that the mobility prior was deliberately chosen to be broad, so that the finite-rate model can approach a quasi-static limit where the transfer is effectively instantaneous.

\begin{figure}[H]
  \centering
  \begin{subfigure}[t]{\linewidth}
    \centering
    \includegraphics[width=\linewidth]{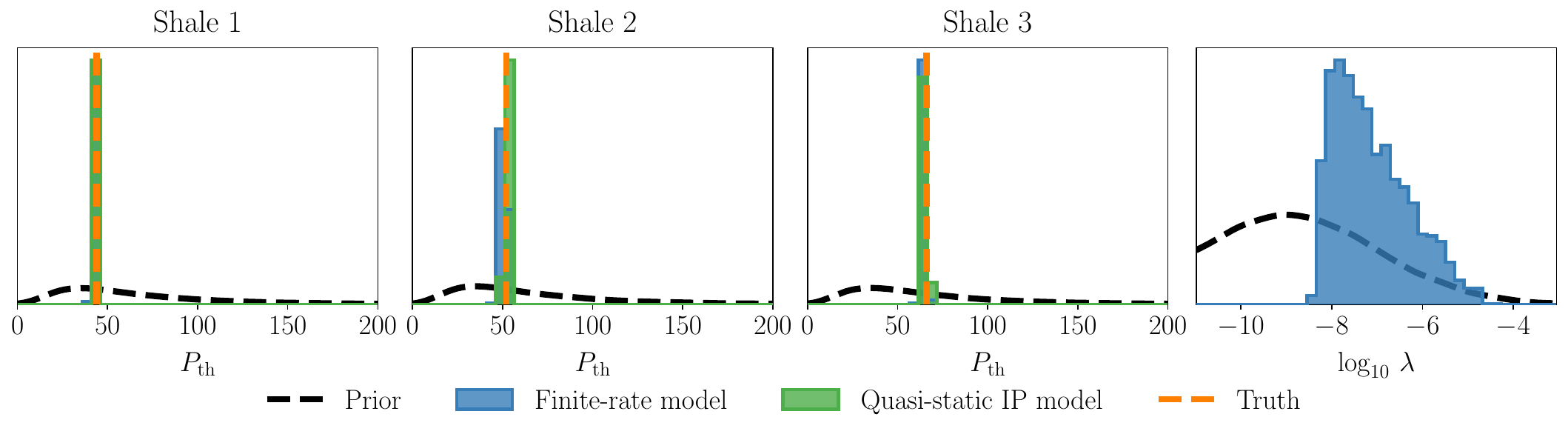}
    \caption{Posterior distributions}
    \label{fig:synth_transfer_quasi_posteriors}
  \end{subfigure}
  \vfill
  \begin{subfigure}[t]{\linewidth}
    \centering
    \includegraphics[width=\linewidth]{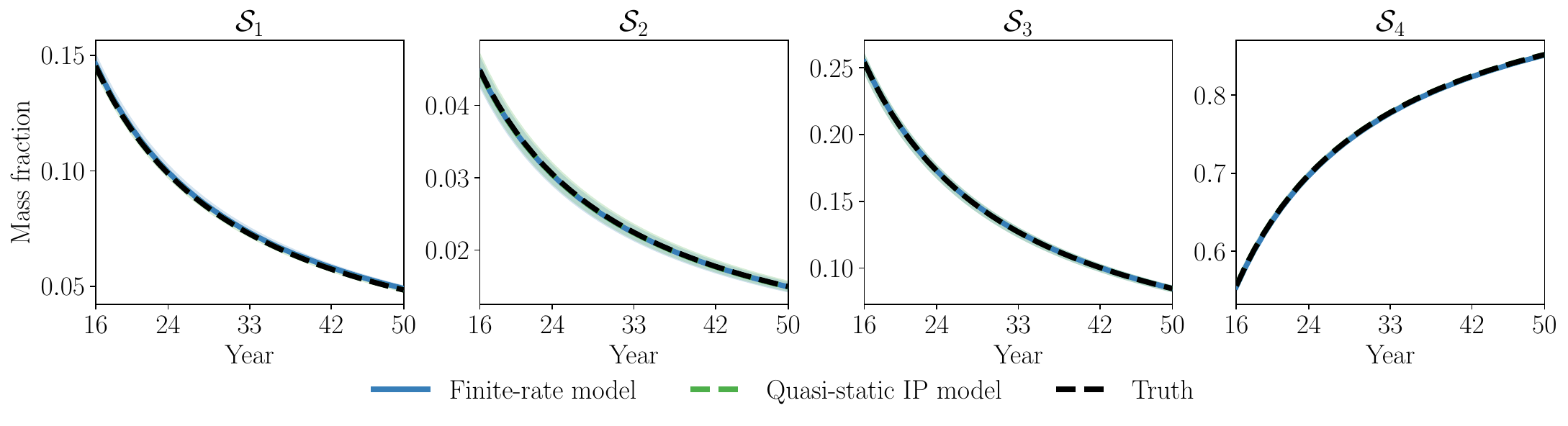}
    \caption{Forecast mass fractions}
    \label{fig:synth_transfer_quasi_forecasts}
  \end{subfigure}

  \caption{Finite-rate and quasi-static fits to the quasi-static synthetic truth. Panel (a) shows the shared entry-pressure priors as dashed black curves and the true entry pressures as dashed orange lines. The quasi-static truth has no true mobility to mark. Panel (b) compares both models with the true layer-wise mass fractions.}
  \label{fig:synth_transfer_quasi}
\end{figure}

\Cref{fig:synth_transfer_quasi_posteriors} shows the posterior distributions for the finite-rate and quasi-static models. The finite-rate model recovers the true entry pressures without apparent bias, while the mobility posterior shifts towards higher values but remains broad. \Cref{fig:synth_transfer_quasi_forecasts} shows that the two models produce nearly identical layer-wise mass forecasts, both closely matching the truth. These results indicate that the finite-rate formulation remains appropriate in the quasi-static limit and can therefore accommodate a broad range of transfer rates.

\subsection{Effect of the plume-detection threshold}\label{seq:plume_detection_extents}

\begin{figure}[H]
  \centering
  \includegraphics[width=0.85\linewidth]{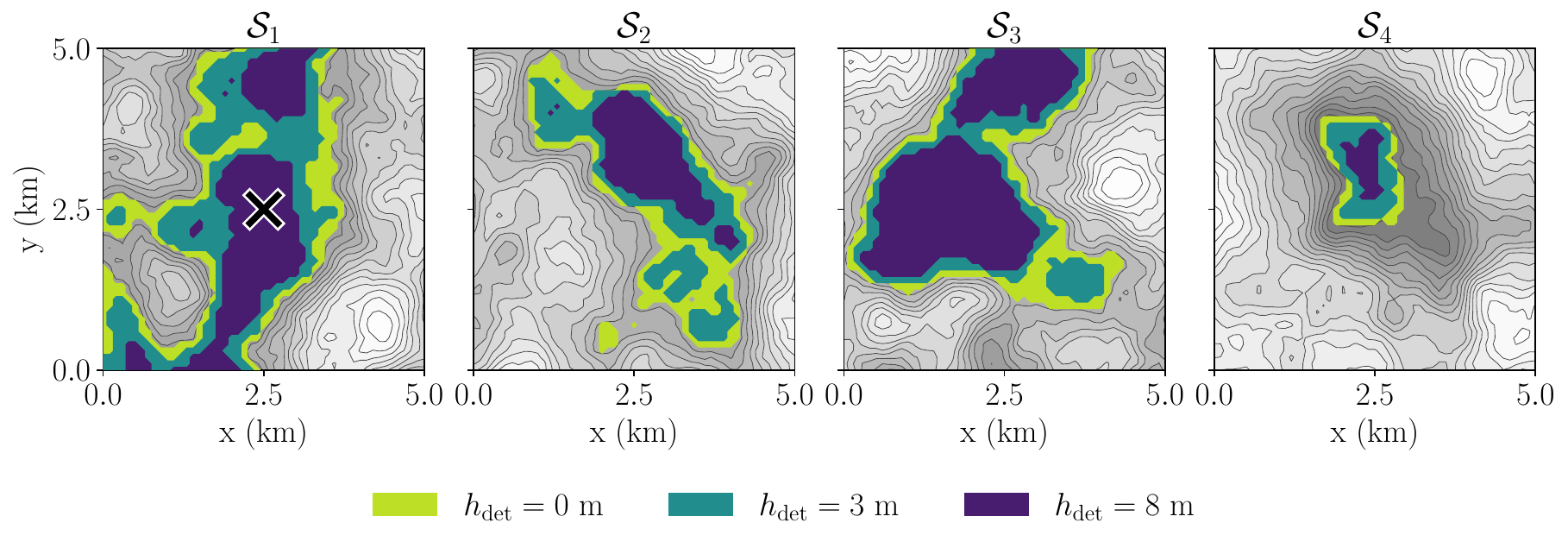}
  \caption{Detected plume extent at year $16$ for detection thresholds of $0$, $3$, and $8\,\mathrm{m}$.}
  \label{fig:synth_detection_extents}
\end{figure}

\Cref{seq:plume_detection} compared the inferred parameters and forecasts for different plume-detection thresholds. \Cref{fig:synth_detection_extents} illustrates how the detected plume extent changes with the threshold. Increasing $h_{\mathrm{det}}$ progressively removes thin plume margins and can eliminate small accumulations from the detected footprint. These geometric changes explain how an incorrect threshold can introduce bias in the inferred entry pressures, which is what was observed in \Cref{fig:synth_transfer_rate_posteriors}. 

\end{appendices}

\backmatter

\bmhead{Acknowledgements}

We acknowledge support from the Centre for Geophysical Forecasting (grant no. 309960).

\section*{Statements and Declarations}

\subsection*{Funding}
This work was supported by the Centre for Geophysical Forecasting (grant no. 309960).

\subsection*{Competing interests}
The authors declare no competing interests.

\subsection*{Data availability}
The code and data is provided in a GitHub repository \url{https://github.com/ellingsvee/CO2IPSimulator}.

\bibliography{bibliography}
\end{document}